\documentclass[11pt]{article}
\pdfoutput=1 
\usepackage{jcappub}
\usepackage{xcolor,float}
\usepackage[utf8]{inputenc}

\def\beq{\beq\begin{aligned}}
\def\eeq{\end{aligned}\eeq}
\newcommand{\gs}{g_\star}
\newcommand{\gss}{g_{\star s}}
\newcommand{\sv}{\langle\sigma v\rangle}
\newcommand{\Tmax}{T_\text{max}}
\newcommand{\amax}{a_\text{max}}
\newcommand{\Trh}{T_\text{RH}}
\newcommand{\arh}{a_\text{RH}}
\newcommand{\Tc}{T_\times}
\newcommand{\ac}{a_\times}
\newcommand{\aini}{a_\text{in}}
\newcommand{\Hini}{H_\text{in}}
\newcommand{\rp}{\rho_\phi}
\newcommand{\rR}{\rho_R}
\newcommand{\Gp}{\Gamma_\phi}
\newcommand{\mphi}{m_\phi}
\newcommand{\Mp}{M_\text{Pl}}
\newcommand{\br}{\text{Br}}

\def\beq{\begin{equation}\begin{aligned}}
\def\eeq{\end{aligned}\end{equation}}

\begin{document}
\begin{flushright}
PI/UAN-2019-654FT
\end{flushright}

\title{\vspace{-25mm} Ultraviolet Freeze-in and\\Non-Standard Cosmologies  }

\author[a,\,b]{Nicol\'as Bernal,}
\author[c]{Fatemeh Elahi,}
\author[d]{\\[5pt]Carlos Maldonado,}
\author[e,\,f]{and James Unwin}
\affiliation[a]{Centro de Investigaciones, Universidad Antonio Nari\~no, Bogot\'a, Colombia}
\affiliation[b]{Institute of High Energy Physics, Austrian Academy of Sciences, Vienna, Austria}
\affiliation[c]{School of Particles and Accelerators,
Institute for Research\\ in Fundamental Sciences (IPM), Tehran, Iran}
\affiliation[d]{Departamento de F\'isica, Universidad de Santiago de Chile, Santiago, Chile}
\affiliation[e]{Department of Physics,  University of Illinois at Chicago, Chicago, Illinois, 60607, USA}
\affiliation[f]{Department of Physics, University of California, Berkeley \& Theoretical Physics Group, LBNL \& Mathematics Sciences Research Institute, Berkeley, CA 94720, USA}
\emailAdd{nicolas.bernal@uan.edu.co}
\emailAdd{felahi@ipm.ir}
\emailAdd{carlos.maldonados@usach.cl}
\emailAdd{~~~~~~~~~~~unwin@uic.edu}

\abstract{
	A notable feature of UV freeze-in is that the relic density is strongly dependent on the highest temperatures of the thermal bath, and a common assumption is that the relevant ``highest temperature'' should be the reheating temperature after inflation $\Trh$. However, the temperature of the thermal bath can be significantly higher in certain scenarios, reaching a value denoted $\Tmax$, a fact which is only apparent away from the instantaneous decay approximation.  Interestingly, it has been shown that if the operators are of sufficiently high mass dimension then the dark matter abundance can be enhanced by a ``boost factor'' depending on $(\Tmax/\Trh)$ relative to naive estimates assuming instantaneous reheating.  We highlight here that in non-standard cosmological histories the critical mass dimension of the operator above at which the instantaneous decay approximation breaks down, and the exponent of the boost factor, depend on the equation of state $\omega$ prior to reheating. We highlight four examples in which the dark matter abundance receives a significant enhancement in the context of gravitino dark matter, the moduli portal, the Higgs portal, and the spin-2 portal (as might arise in bimetric gravity models). We comment on the transition from kination domination to radiation domination as a motivated example of non-standard cosmologies.}

\maketitle

\section{Introduction}
\vspace{-1mm}

The expected signals and constraints on dark matter (DM) are dictated by its interactions with the states of the Standard Model, which in turn are informed by the cosmological evolution of abundances which establishes the DM relic density. In this work we consider the scenario in which the relic abundance of DM is set via the DM freeze-in production mechanism \cite{Hall:2009bx} (for a recent review see \cite{Bernal:2017kxu}). More specifically, here we focus on the subcase of Ultraviolet (UV) freeze-in~\cite{Elahi:2014fsa} for which the temperature of the thermal bath is always lower that the mass of the mediator states which connect the DM to the Standard Model implying that the connector operators are non-renomalisable with mass dimension $5+n/2$ (for $n$ even,\footnote{We parameterise the mass dimension of operators in this somewhat odd fashion in order to match the conventions of earlier papers~\cite{Garcia:2017tuj,Banerjee:2019asa,Bernal:2018qlk,Chowdhury:2018tzw,Chen:2017kvz} and such that the cross section can be expressed as $\sv\propto T^{n}$.} with $n\geq0$).
Thus in UV freeze-in the production  cross section of DM from interactions in the Standard Model thermal bath are of the form
\beq
\sv\sim\frac{T^{n}}{\Lambda^{2+n}}~,
\label{cross-section}
\eeq
where $T$ is the bath temperature and $\Lambda$ is a dimensional quantity which is parametrically the mass scale of the states which mediate interactions between the DM and the Standard Model. 
Recall from above that $n=0$ corresponds to mass dimension 5 in which case $\sv$ is temperature independent, and $n=2$ corresponds dimension 6 operators for which $\sv\propto T^2$.

It follows that the DM abundance due to eq.~(\ref{cross-section}) is expected to be of the form~\cite{Elahi:2014fsa}
\beq
Y\sim \int^{\Trh}_0\frac{\Mp\,T^{n}}{\Lambda^{n+2}}\sim\frac{\Mp\,\Trh^{n+1}}{\Lambda^{n+2}}~.
\label{step}
\eeq
This integral is cutoff at some temperature which is the highest temperature of the radiation bath, which we take here to be the reheating temperature $\Trh$ following inflation, and the final abundance is highly sensitive to this cutoff.

More generally, if the early universe is dominated by some other energy density which subsequently decays (such as an early period of matter domination) then the restoration of radiation domination reheats the thermal bath altering the temperature evolution and the reheating temperature at which decays are complete $\Trh$ can be decoupled from the physics of inflation.  Indeed, DM in the context of non-standard cosmology has recently gained increasing interest, see e.g.~\cite{Davoudiasl:2015vba, Randall:2015xza, Tenkanen:2016jic, Dror:2016rxc, DEramo:2017gpl,  Hamdan:2017psw, Visinelli:2017qga, Drees:2017iod, DEramo:2017ecx, Maity:2018dgy, Bernal:2018kcw, Arbey:2018uho, Drees:2018dsj, Betancur:2018xtj, Maldonado:2019qmp, Poulin:2019omz, Tenkanen:2019cik, Arias:2019uol, Bramante:2017obj,Bernal:2018ins,DiMarco:2018bnw}.  Throughout we will use $\Trh$ to indicate the temperature of the thermal bath following the final reheating event prior to the onset of standard cosmology (which could be simply due to inflaton decays) and for ease of writing, we shall often discuss the physics in terms of inflaton decay. Notably, observational constraints from Big Bang Nucleosynthesis require $\Trh\gtrsim10$ MeV~\cite{Sarkar:1995dd}.

Importantly, during reheating in which the Standard Model thermal bath is produced and the universe transitions to radiation domination, the bath temperature may rise to a value $T_{\rm max}$ which exceeds $\Trh$ \cite{Giudice:2000ex}. That the maximum temperature of the thermal bath may reach $T_{\rm max}>\Trh$ prior to cooling is not apparent if one takes the instantaneous decay approximation for reheating. It is quite plausible that the DM relic density may be established during this reheating period, set by its production or annihilation cross section, in which case the DM abundance will significantly differ from freeze-in or freeze-out calculations assuming radiation domination, see e.g.~\cite{Chung:1998rq,McDonald:1989jd,Giudice:2000ex,Gelmini:2006pw,Allahverdi:2002pu,Allahverdi:2002nb}. 
In particular, it has been observed by Garcia-Mambrini-Olive-Peloso \cite{Garcia:2017tuj}  that if the DM is produced during the transition from matter to radiation domination  via an effective operator which connects the DM and Standard Model states leading to a cross section of the form of eq.~\eqref{cross-section},  for $n> 6$  the DM abundance is  enhanced by a ``boost factor''  $B\sim(\Tmax/\Trh)^{n-6}$. Whereas for $n\leq6$ the difference between the standard UV freeze-in calculations~\cite{Elahi:2014fsa,Hall:2009bx}, which assume an instantaneous transition  (i.e.~employs the instantaneous decay approximation), differ only by an $\mathcal{O}(1)$ factor from calculations taking into account non-instantaneous reheating \cite{Garcia:2017tuj}. Subsequent papers~\cite{Banerjee:2019asa,Bernal:2018qlk,Chowdhury:2018tzw,Chen:2017kvz, Bhattacharyya:2018evo, Kaneta:2019zgw} have explored the impact of this boost factor in specific models. 

In this work we demonstrate that the critical value $n_c$ for which the DM relic abundance is enhanced when taking into account non-instantaneous reheating depends on the equation of state $\omega$ prior to reheating. For an early period of matter domination ($\omega=0$) the critical value is $n_c=6$, and the details of the process of reheating are important for operators with mass dimension 8 and higher. More generally, if the early universe features a period of non standard cosmology in which the universe is neither radiation or matter dominated, but rather the dominant energy density evolves as $a^{-3(\omega+1)}$ then $n_c$ will depend on $\omega$, as will the exponent of the boost factor that enhances the DM abundance relative to the sudden decay approximation. Thus different cosmological assumptions can potentially lead to significant enhancements of the DM relic density due to UV freeze-in. Additionally, it was recently highlighted that the initial value of $\omega$ impacts the DM relic density if it is established due to freeze-out or renormalisable (IR) freeze-in during the era of particle decays leading to the transition to radiation domination~\cite{Bernal:2018kcw, Maldonado:2019qmp}.

This paper is structured as follows: In Section \ref{S2} we outline the general framework in a model independent approach, in particular we derive the dependence of the critical dimension $n_c$ and the boost factor $B$ on the equation of state prior to reheating $\omega$ for general operator dimension parameterised by $n$. We subsequently explore how the DM abundance varies for different choices of $\omega$ and $n$. In Section \ref{S3} we highlight four specific examples in which the DM abundance receives a significant enhancement, specifically we examine gravitino DM  in High Scale Supersymmetry \cite{Garcia:2017tuj,Hall:2009nd}, the moduli portal  \cite{Chowdhury:2018tzw}, the vector Higgs portal, and the massive spin-2 portal~\cite{Bernal:2018qlk}. In Section \ref{S4} we provide a summary and some concluding remarks.


\section{General framework}
\label{S2}

We first take a model independent approach to outline the impact of non-instantaneous reheating on UV freeze-in assuming a general initial equation of state $\omega$ for the early universe.

\subsection{UV freeze-in in the sudden decay approximation}

The evolution of the DM number density $n$ is given by the Boltzmann equation
\beq\label{eq:cosmo1}
\frac{dn}{dt}+3\,H\,n=-\sv\left(n^2-n_\text{eq}^2\right),
\eeq
where $n_\text{eq}=a\,g\,\pi^{-2}\zeta(3) \,T^3$ is the equilibrium DM number density for relativistic particles, in terms of  $g$ the DM number of degrees of freedom, and the constant $a$ which is $a=1$ for bosonic DM and $a=3/4$ for a fermion,  but henceforth we simply set $a$ to 1. 
Additionally, here $H=\left[\rR/(3\Mp^2)\right]^{1/2}$ is the Hubble expansion rate expressed in terms of the reduced Planck mass $\Mp$ and the Standard Model energy density $\rR(T)\equiv\frac{\pi^2}{30}\,\gs(T)\,T^4$.

In the sudden decay approximation for reheating the Standard Model entropy density is always conserved, thus one can rewrite eq.~\eqref{eq:cosmo1} as a function of the dimensionless variable $Y\equiv n/s$ as follows
\beq\label{eq:cosmo1b}
	\frac{dY}{dT}=\frac{\sv\,s}{H\,T}\left(Y^2-Y_\text{eq}^2\right),
\eeq
where 	$s(T)=\frac{2\pi^2}{45}\,\gss(T)\,T^3$ is the entropy density for $\gs(T)$ and $\gss(T)$ the effective numbers of relativistic degrees of freedom of the Standard Model radiation and entropy densities.

As preempted in eq.~(\ref{cross-section})  we take the thermally averaged DM production cross section to be a function of the thermal bath temperature $T$
\beq
	\sv=\frac{T^n}{\Lambda^{n+2}}\,.
\eeq
The scale $\Lambda$ corresponds to the cutoff of the effective field theory and typically corresponds to the mass of some mediator that connects the Standard Model and DM. For this  effective operator description to be valid $\Lambda$ must be the highest scale in the calculation, and we assume throughout the hierarchy $m<T\ll\Lambda$, where $m$ is the DM mass.
If the DM abundance is initially negligible and the production cross section is sufficient small that DM remains out of chemical equilibrium with the Standard Model bath and within the regime that the sudden decay approximation for the inflaton is valid, then eq.~\eqref{eq:cosmo1b} has an analytical solution given~by
\beq\label{eq:Y0T}
	Y(T)=\frac{135\,\zeta(3)^2}{2\pi^7\,(n+1)}\sqrt{\frac{10}{\gs}}\frac{g^2}{\gss}\frac{\Mp}{\Lambda^{n+2}}\left[\Trh^{n+1}-T^{n+1}\right],
\eeq
where, assuming instantaneous decays, $\Trh$ corresponds to the temperature at which the inflaton decays, and therefore to the maximal temperature reached by the Standard Model thermal bath. This enters as the upper limit of the temperature integral leading to eq.~(\ref{eq:Y0T}), as indicated in eq.~(\ref{step}).
Furthermore, the asymptotic value $Y_\infty$ for $T\ll\Trh$ then gives the DM relic abundance which is found to be 
\beq\label{eq:Y0}
	Y_\infty=\frac{135\,\zeta(3)^2}{2\pi^7\,(n+1)}\sqrt{\frac{10}{\gs}}\frac{g^2}{\gss}\frac{\Mp}{\Lambda^{n+2}}\Trh^{n+1}~,
\eeq
where we have neglected the small deviation due to temperature evolution of the numbers of relativistic degrees of freedom. Notably, the majority of the DM is produced near the highest temperatures ($T\sim\Trh$) reached by the universe, which is characteristic of UV freeze-in.


\subsection{Reheating in non-standard cosmologies}

While reheating is commonly approximated as an instantaneous event, the decay of the inflaton into Standard Model radiation is a continuous process, reasonably characterised by an exponential decay law \cite{Scherrer:1984fd}. The evolution of the Standard Model and inflaton $\phi$ abundances can be tracked via a pair of Boltzmann equations; the  $\phi$ energy density follows  
\beq
\frac{d\rp}{dt}+3(1+\omega)\,H\,\rp&=-\Gp\,\rp\,,
\label{eq:cosmo2}
\eeq
where $\omega\equiv p_\phi/\rp$ corresponds to the equation of state of $\phi$, with $p_\phi$ and $\rp$ being the $\phi$ pressure and  energy density, and
 $\Gp$ is the total decay width of $\phi$. The Hubble expansion rate $H$ receive contributions from all species, i.e.~both the Standard Model and $\phi$, thus 
    $H^2=(\rp+\rR)/(3\,\Mp^2)$. The evolution of the Standard Model bath evolves according to
\beq
	\frac{ds}{dt}+3\,H\,s&=\frac{\gss}{\gs}\frac{\Gp\,\rp}{T}\,.
\label{eq:cosmo3}
\eeq
Using the form $s(T)=\frac{2\pi^2}{45}\,\gss(T)\,T^3$  with  eq.~\eqref{eq:cosmo3} gives the temperature evolution of the Standard Model thermal bath as a function of the scale factor $a$
\beq\label{eq:cosmo3b}
    \frac{dT}{da}=\left(1+\frac{T}{3\,\gss}\frac{d\gss}{dT}\right)^{-1}\left[\frac{\gss}{\gs}\frac{\Gamma_\phi\,\rp}{3\,H\,s\,a}-\frac{T}{a}\right].
\eeq
Moreover, if the variation of relativistic degrees of freedom can be neglected (as is typically very reasonable), eq.~\eqref{eq:cosmo3} is usually rewritten in terms of the Standard Model energy density
\beq
	\frac{d\rR}{dt}+4\,H\,\rR=+\Gp\,\rp~.
\eeq
Since the  $\Trh$ is defined as the temperature at which the equality $H(T=\Trh)=\Gp$ holds,
the total decay width $\Gp$  can  be expressed as a function of $\Trh$ as
\beq\label{eq:gamma}
	\Gp=\frac{\pi}{3}\sqrt{\frac{\gs(\Trh)}{10}}\frac{\Trh^2}{\Mp}~.
\eeq

Ignoring the variation of the number of relativistic degrees of freedom $\gs$ and $\gss$, eqs.~\eqref{eq:cosmo2} and \eqref{eq:cosmo3} can be analytically solved. Prior to reheating the $\phi$ energy density, which dominates the energy density of the universe, evolves as
\beq
	\rp(a)=\rp(\aini)\left[\frac{\aini}{a}\right]^{3(1+\omega)}=3\,\Mp^2\,\Hini^2\,\left[\frac{\aini}{a}\right]^{3(1+\omega)},
	\eeq
	and the radiation energy density evolves according to (for $\omega\neq5/3$)\footnote{The case $\omega=5/3$ must be treated separately since integrating leads to a logarithm  of the form $\ln(a/\aini)$ rather than the factor $(a^\frac{5-3\omega}{2}-\aini^\frac{5-3\omega}{2})$. Simply for brevity, we suppress this special case.}
\beq
	\rR(a)=
		\frac{6}{5-3\omega}\Mp^2\,\Hini\,\Gp\,\frac{\aini^{\frac32(1+\omega)}}{a^4}\left[a^\frac{5-3\omega}{2}-\aini^\frac{5-3\omega}{2}\right]~,
		\label{eq:rR}
\eeq
where $a=\aini$ is the scale factor at some arbitrary initial point, we  take the initial condition $\rR(\aini)=0$ and thus
  $\Hini\equiv H(\aini)=\sqrt{\rp(\aini)/(3\Mp^2)}$.

It follows that the  thermal bath reaches a maximum temperature $\Tmax$ when only a small fraction of the inflaton has decayed~\cite{Chung:1998rq, Giudice:2000ex}, with $\Tmax$ corresponding to the scale factor 
\beq
	\amax=
		\aini\left[\frac{8}{3(1+\omega)}\right]^\frac{2}{5-3\omega}~.
\label{amax}
\eeq
Note that $a_{\rm max}$ depends on $\omega$, as does the evolution of the bath temperature \cite{Maldonado:2019qmp}. If $\Tmax$ is taken as an input parameter, the corresponding initial energy density in $\phi$  is given by
\beq
	\rp(\aini)=
		\frac13\left(\frac{\gs\,\pi^2}{20}(1+\omega)\left[\frac{8}{3(1+\omega)}\right]^\frac{8}{5-3\omega}\frac{\Tmax^4}{\Mp\,\Gp}\right)^2~,
\eeq
which implies an initial Hubble expansion rate of the form
\beq
	\Hini=
		\frac{\gs\,\pi^2}{60}(1+\omega)\left[\frac{8}{3(1+\omega)}\right]^\frac{8}{5-3\omega}\frac{\Tmax^4}{\Mp^2\,\Gp}~.
\label{Hini}
\eeq

\begin{figure}[t!]
\begin{center}
\includegraphics[height=0.3\textwidth]{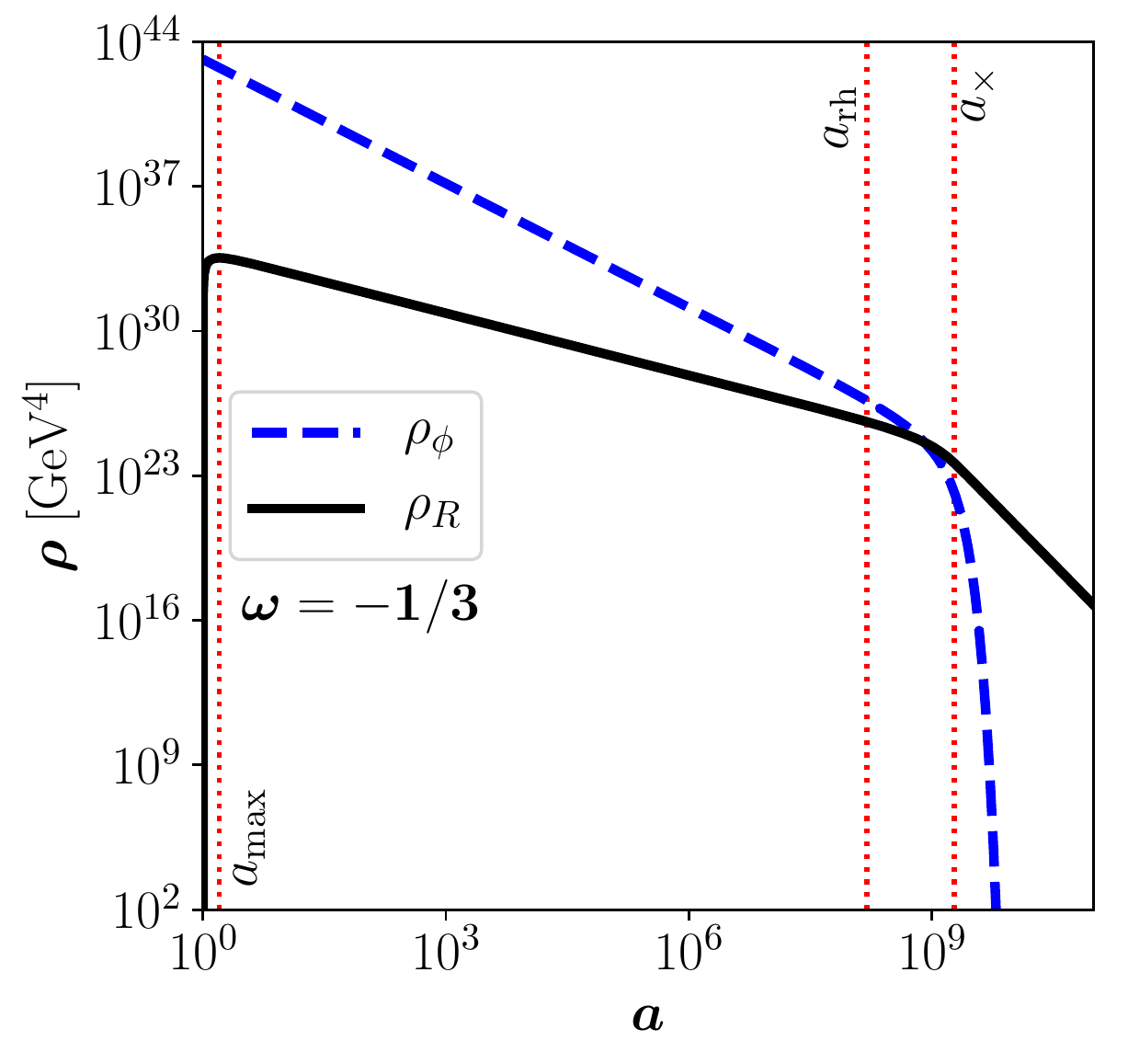}
\includegraphics[height=0.3\textwidth]{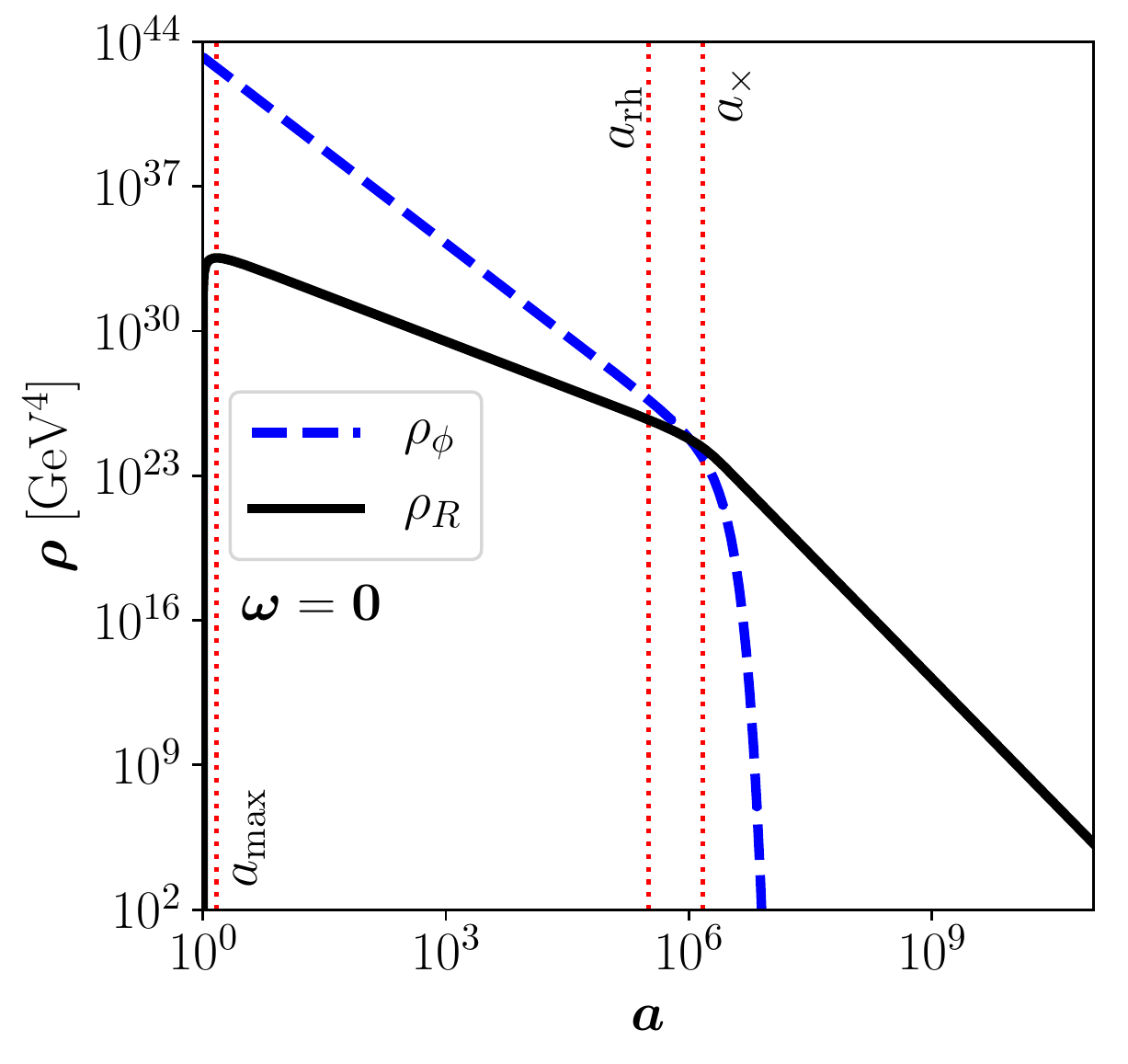}
\includegraphics[height=0.3\textwidth]{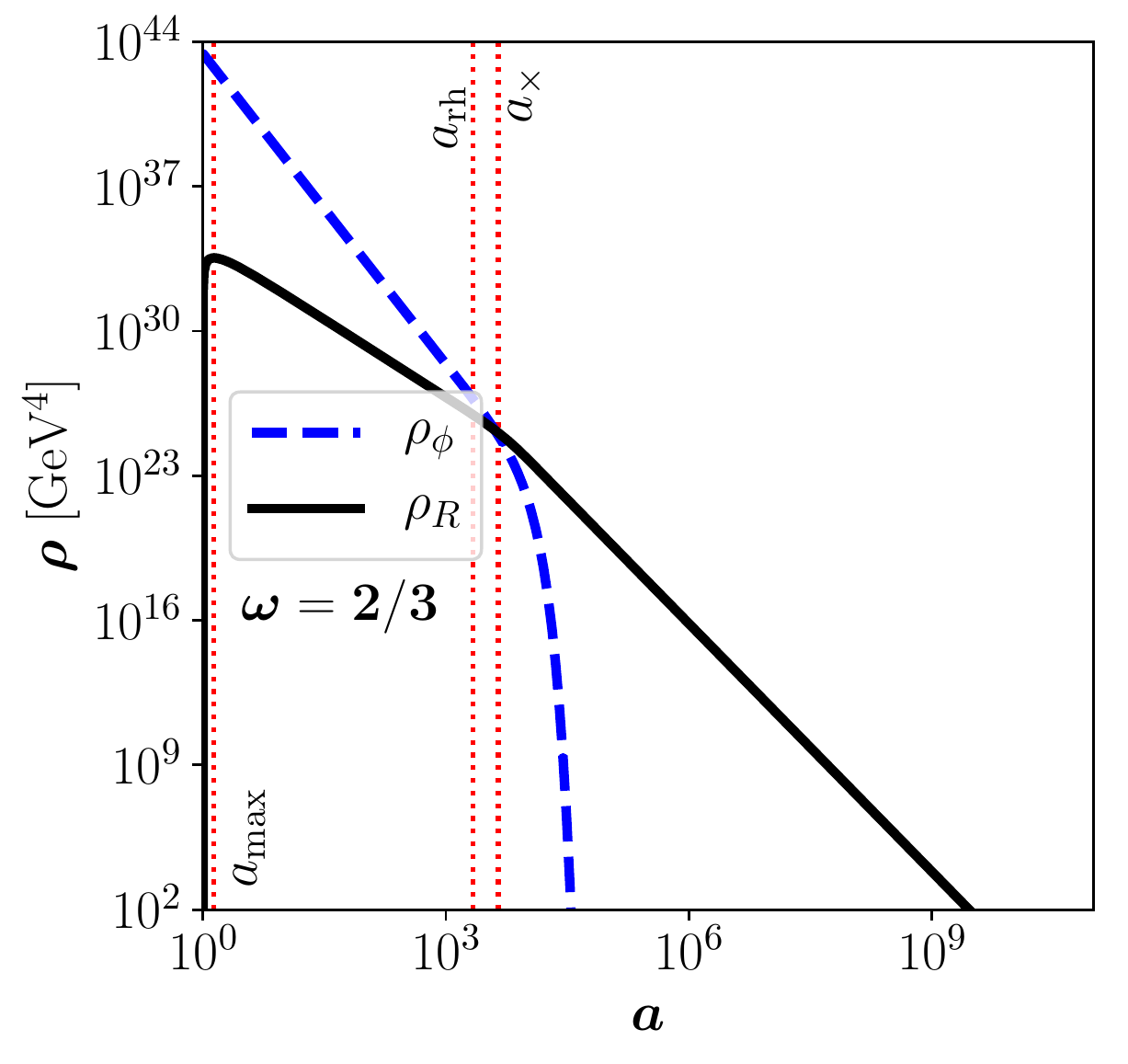}\\
\includegraphics[height=0.31\textwidth]{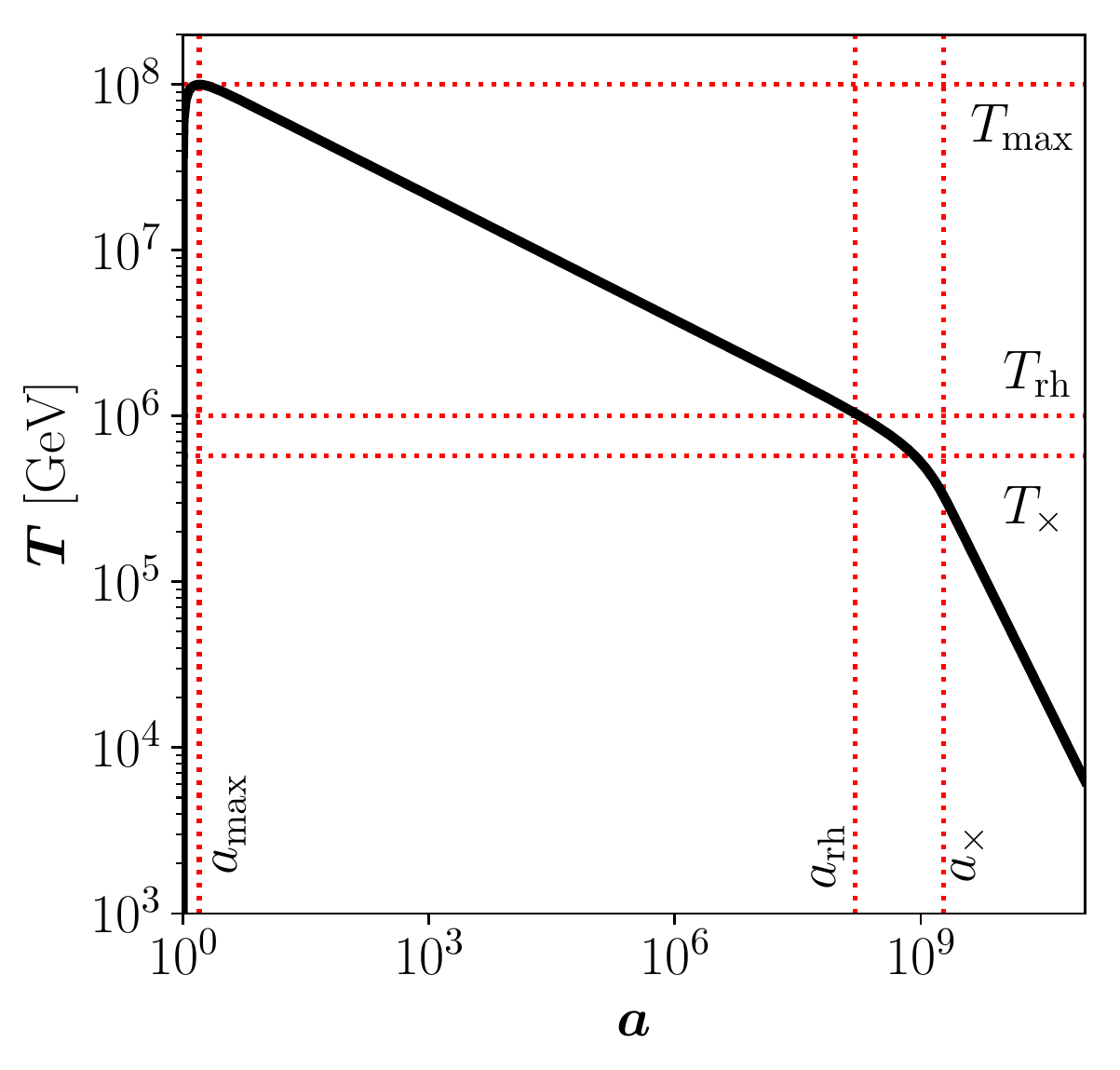}
\includegraphics[height=0.31\textwidth]{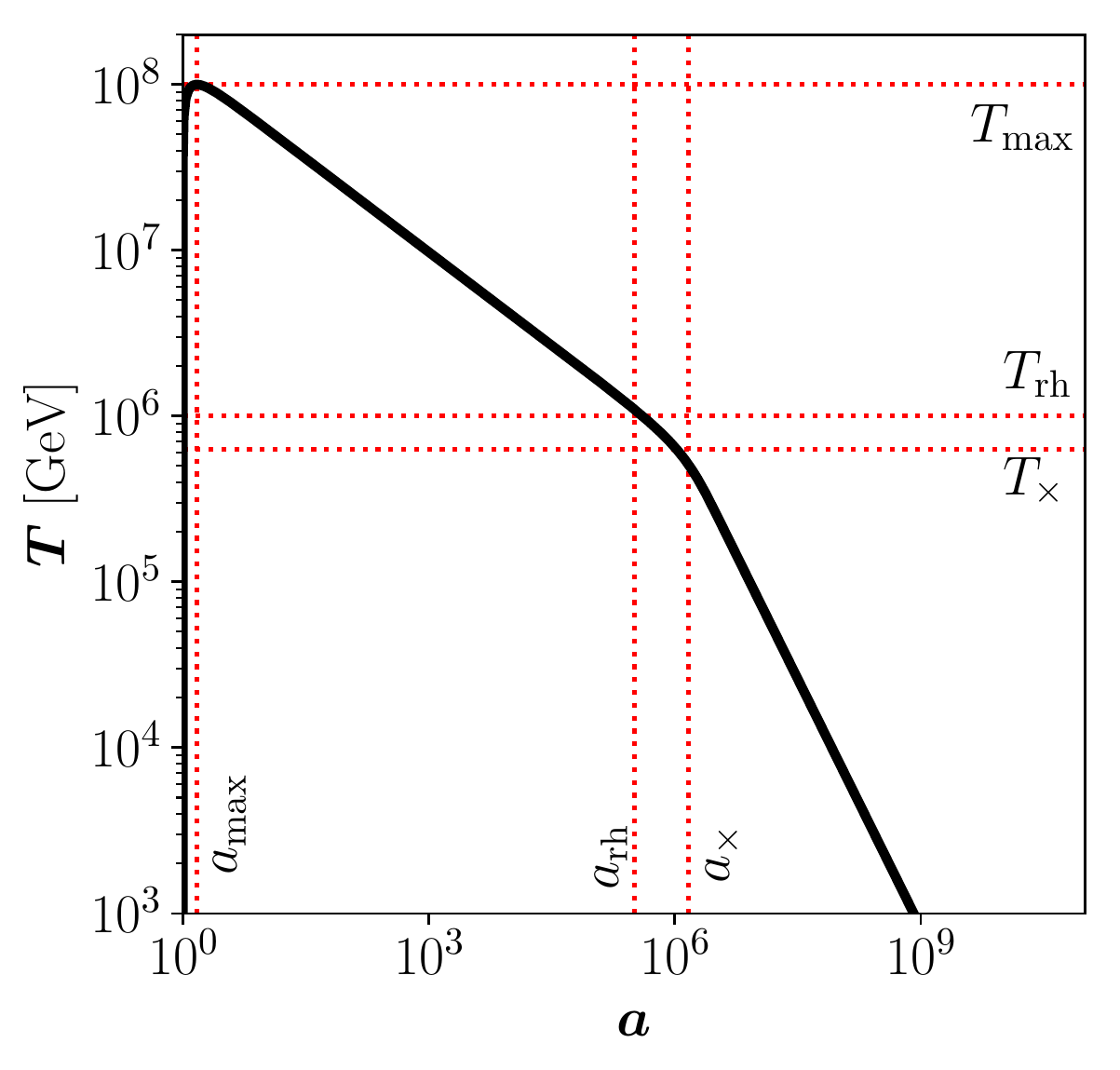}
\includegraphics[height=0.31\textwidth]{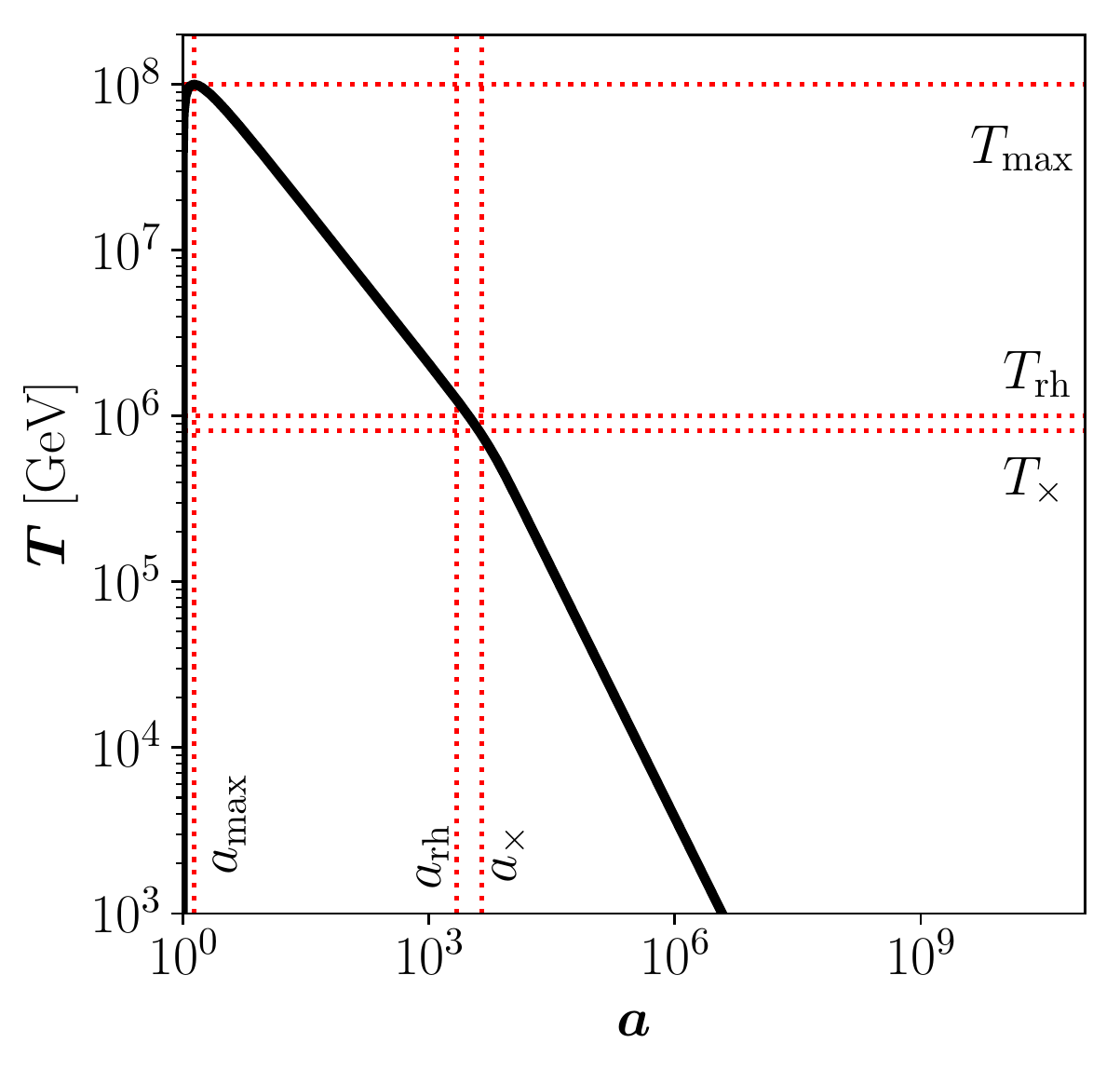}
	\vspace{-5mm}
	\caption{Evolution of the energy densities (upper panels) and Standard Model thermal bath temperature (lower panels)
	 for $\Trh=10^6$~GeV and $\Tmax=10^8$~GeV.
	The left panels depict the case $\omega=-1/3$, central panels $\omega=0$ and right panels $\omega=2/3$.
	The dotted lines corresponding to $a=\amax$, $\arh$ and $\ac$ (from left to right) and $T=\Tmax$, $\Trh$ and $\Tc$ (from top to bottom) are overlaid.  
	}\label{fig:evolution2}
	\vspace{-5mm}
\end{center}
\end{figure}

Moreover, from eq.~\eqref{eq:rR} it follows  that the radiation energy density scales like
\beq
    \rR(a)\propto 
    \begin{cases}
	    a^{-\frac32(1+\omega)} & \text{ for }\amax\ll a\ll\ac,\\[12pt]
	    a^{-4} & \text{ for }\ac\ll a\,,
    \end{cases}
    \label{rpa}
\eeq
where $\ac$ is the point at which $\rp=\rR$ (assuming they can be treated as independent power laws).
Figure~\ref{fig:evolution2}  shows the evolution of energy densities and the bath temperature for $\omega=-1/3$ (left panels), 0 (right panels) and 2/3 (right panels).


Equation \eqref{rpa} implies that $ T(a)\propto a^{-\frac38(1+\omega)}$ for $\amax\ll a\ll\ac$ and  during radiation domination $T(a)\propto a^{-1}$. Moreover, in terms of temperature the point of $\phi$-radiation equality $\ac$ corresponds to a thermal bath temperature of $\Tc\equiv T(\ac)$  given by
\beq\label{eq:Teq}
	\Tc^4=\begin{cases}
		\frac{4}{(5-3\omega)^2}\,\Trh^4&\text{ for }\omega<\frac53,\\[12pt]
		\frac{180\Mp^2}{\pi^2\,\gs(\Tc)}\left[16\left(\frac{1}{3\omega-5}\Gp\right)^{3(1+\omega)}\Hini^{3\omega-5}\right]^\frac{1}{3\omega-1}&\text{ for }\omega>\frac53,
	\end{cases}
\eeq
and the corresponding scale factors for $\Tc$ and $\Trh$ are
\beq
	\ac&=&\begin{cases}
		\aini\left[\frac{5-3\omega}{2}\frac{\Hini}{\Gp}\right]^\frac{2}{3(1+\omega)}&\text{ for }\omega<\frac53,\\[12pt]
		\aini\left[\frac{3\omega-5}{2}\frac{\Hini}{\Gp}\right]^\frac{1}{3\omega-1}&\text{ for }\omega>\frac53,
	\end{cases}
	\eeq
where	$\arh=\ac(\Tc/\Trh)^\frac{8}{3(1+\omega)}$.
Note that $\ac$ and $\Tc$ imply breaks in the power law scaling, as can be seen also in Figure \ref{fig:evolution2}.

\newpage


\subsection{UV freeze-in in non-standard cosmologies}

\label{general-nst}
We next study DM production via UV freeze-in away from the sudden decay approximation and where we assume a general equation of state $\omega$ for the period preceding the decays of $\phi$ to Standard Model states.\footnote{Equations~\eqref{eq:cosmo1} and~\eqref{eq:cosmo3} could be generalized to include possible direct decays of $\phi$ into DM \cite{Gelmini:2006pw,Drees:2017iod}, and non-instantaneous thermalisation of the Standard Model bath \cite{Harigaya:2014waa,Mukaida:2015ria,Harigaya:2019tzu}, however we neglect both effects here.}  As is well known, if the universe is matter or radiation dominated these components redshift as $a^{-3}$ or $a^{-4}$ respectively and correspondingly this implies $\omega=0$ or $\omega=1/3$. More generally $\omega$ can take a range of values and accordingly the comoving Hubble volume evolves as $(aH)^{-1}\propto a^{(1+3\omega)/2}$. Indeed, for a real scalar field $\phi$ with a positive potential which dominates the energy density of the early universe, then the equation of state can take values in $\omega\in(-1,1)$ and allowing for negative potentials then higher values of $\omega$ can be realised \cite{DEramo:2017gpl}. While scenarios of $\omega>1$ are less common, they can arise in models with scalars with periodic  potentials \cite{Gardner:2004in,Choi:1999xn}, scalar-tensor models \cite{Dutta:2016htz}, and brane world cosmology  \cite{Okada:2004nc,Meehan:2014bya}.

We now solve the system of coupled Boltzmann equations~\eqref{eq:cosmo1}, \eqref{eq:cosmo2}, and \eqref{eq:cosmo3}.
To track the evolution of particle populations during the era of reheating in which the Standard Model entropy is not conserved due to the decays of $\phi$, it is better to rewrite eq.~\eqref{eq:cosmo1} in terms of the comoving number density $N\equiv n\times a^3$  as follows
\beq\label{eq:cosmo1a}
	\frac{dN}{da}=-\frac{\sv}{a^4\,H}\left(N^2-N_\text{eq}^2\right).
\eeq

Then for $\Tmax\geq T\geq\Tc$, eq.~\eqref{eq:cosmo1a} admits the analytical solution 
\beq\label{eq:N1}
	N(T)=\frac{8\,\zeta(3)^2\,g^2}{3\pi^4\,(n-n_c)(1+\omega)}\left[\frac{\ac^{3+\omega}}{\aini^{1+\omega}}\right]^\frac32\frac{\Tc^{4\frac{3+\omega}{1+\omega}}}{\Lambda^{n+2}\,\Hini}\left[\Tmax^{n-n_c}-T^{n-n_c}\right],
\eeq
where $n_c$ indicates the critical value for $n$, given by
\beq\label{eq:nc}
	n_c\equiv 2\times \left(\frac{3-\omega}{1+\omega}\right).
\eeq

The $\omega$-dependent quantity $n_c$ denotes the critical threshold, for which  DM freeze-in via an operator of mass dimension $n > n_c$ will be parametrically enhanced.
In Table~\ref{tab:critical} we highlight the values for $n_c$ that arise for certain values of $\omega$.
\begin{table}[H]
	\centering
\begin{tabular}{|c||c|}
	\hline
	$\boldsymbol{\omega}$ & ~~$\boldsymbol{n_c}$~~ \\
	\hline\hline
	-1/3 & 10 \\
	-1/5 & 8 \\
	0 (matter) & 6 \\
	1/3 (radiation-like)~~ & 4\\
	1 (kination) & 2 \\
	\hline
\end{tabular}
	\caption{Critical values for $n$ for different values of $\omega$.}
	\label{tab:critical}
\end{table}
\noindent
We highlight that the case $\omega=1$ in Table~\ref{tab:critical} corresponds to `kination domination'~\cite{Spokoiny:1993kt},  implying that the kinetic energy of a scalar field (the $\dot\phi$ term)  dominates the energy density of the universe. This is a concrete scenario of non-standard cosmology  and, indeed, occurs in certain models of inflation.
We also note that for $\omega\rightarrow-1$ then $n_c\rightarrow\infty$, with the limiting case $\omega=-1$ corresponding to dark energy (or quintessence~\cite{Ratra:1987rm,Caldwell:1997ii}) domination, however the analysis we present breaks down in this limit.

Even though the Standard Model entropy density is not conserved when $\phi$ is decaying, the DM yield $Y$ can be defined from~\eqref{eq:N1} as follows
\beq\label{eq:convert}
	Y(T)=\frac{N(T)}{s(T)\,a^3}=\frac{45}{2\pi^2\,\gss}\frac{N(T)}{\ac^3}\left[\frac{T^{5-3\omega}}{\Tc^8}\right]^\frac{1}{1+\omega},
\eeq
and its asymptotic limit can be estimated by taking $T\to\Tc$
\beq\label{eq:Yrh}
	Y(\Tc)=\frac{180\,\zeta(3)^2\,g^2}{\pi^7\,\gss}\sqrt{\frac{10}{\gs}}\frac{1}{(n-n_c)(1+\omega)}\frac{\Mp\,\Tc^{\frac{7-\omega}{1+\omega}}}{\Lambda^{n+2}}\left[\Tmax^{n-n_c}-\Tc^{n-n_c}\right].
\eeq
Let us note that eqs.~\eqref{eq:N1} and~\eqref{eq:Yrh} are only valid for $n\ne n_c$.
In the  case that $n=n_c$ these expressions are modified as follows
\beq\label{eq:N2}
	N(T)=\frac{\zeta(3)^2\,(2+n)\,g^2}{3\pi^4}\left[\frac{\ac^{6+n}}{\aini^4}\right]^\frac{3}{2+n}\frac{\Tc^{6+n}}{\Lambda^{2+n}\,\Hini}\,\ln\frac{\Tmax}{T},
\eeq
and the asymptotic limit for the DM yield is
\beq\label{eq:Yrhcrit}
	Y(\Tc)=\frac{45\,\zeta(3)^2\,(n+2)\,g^2}{2\pi^7\,\gss}\sqrt{\frac{10}{\gs}}\frac{\Mp\,\Tc^{1+n}}{\Lambda^{2+n}}\ln\frac{\Tmax}{\Tc}.
\eeq

\begin{figure}[t!]
\centerline{
\includegraphics[height=0.34\textwidth]{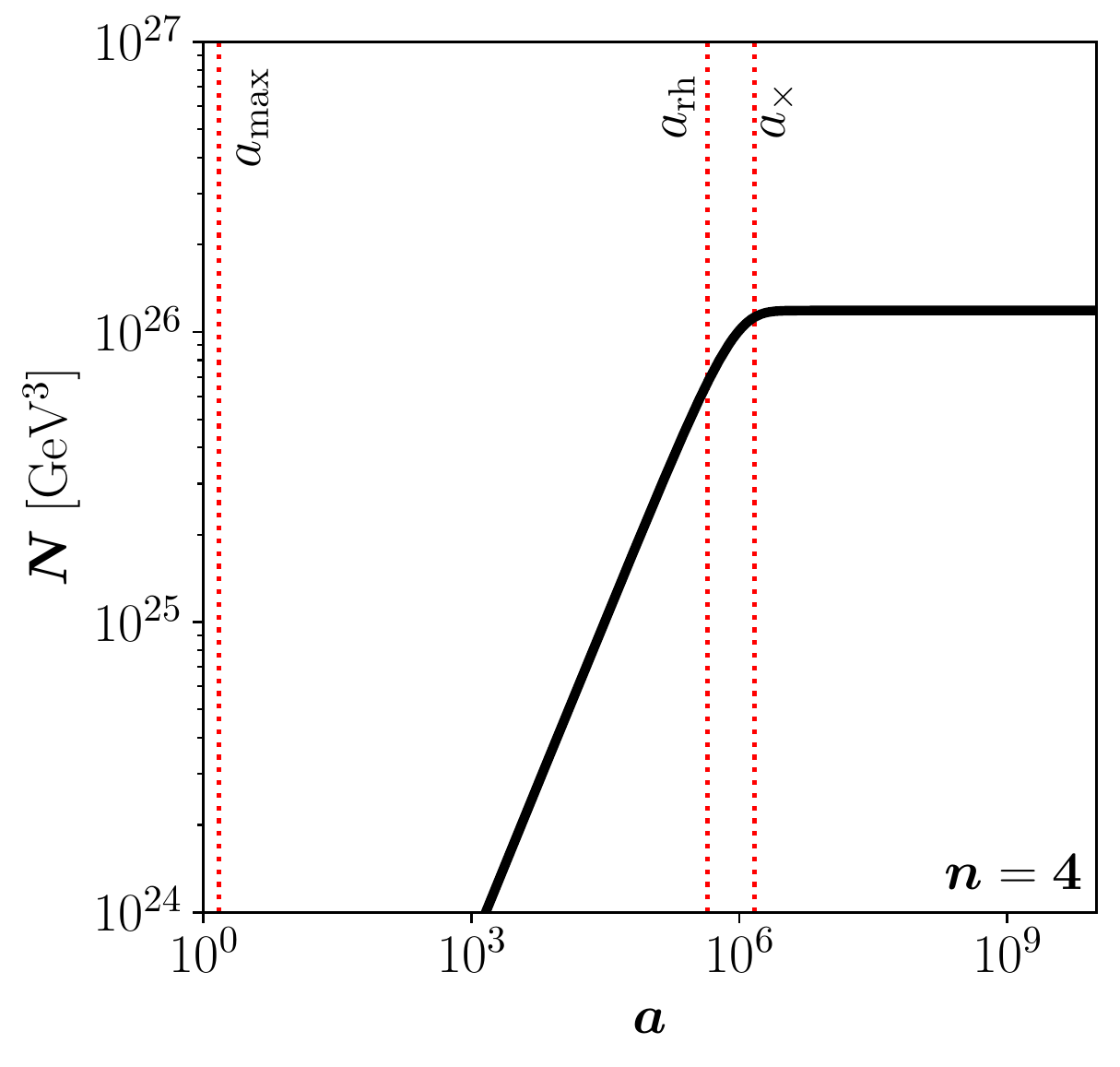}
\includegraphics[height=0.34\textwidth]{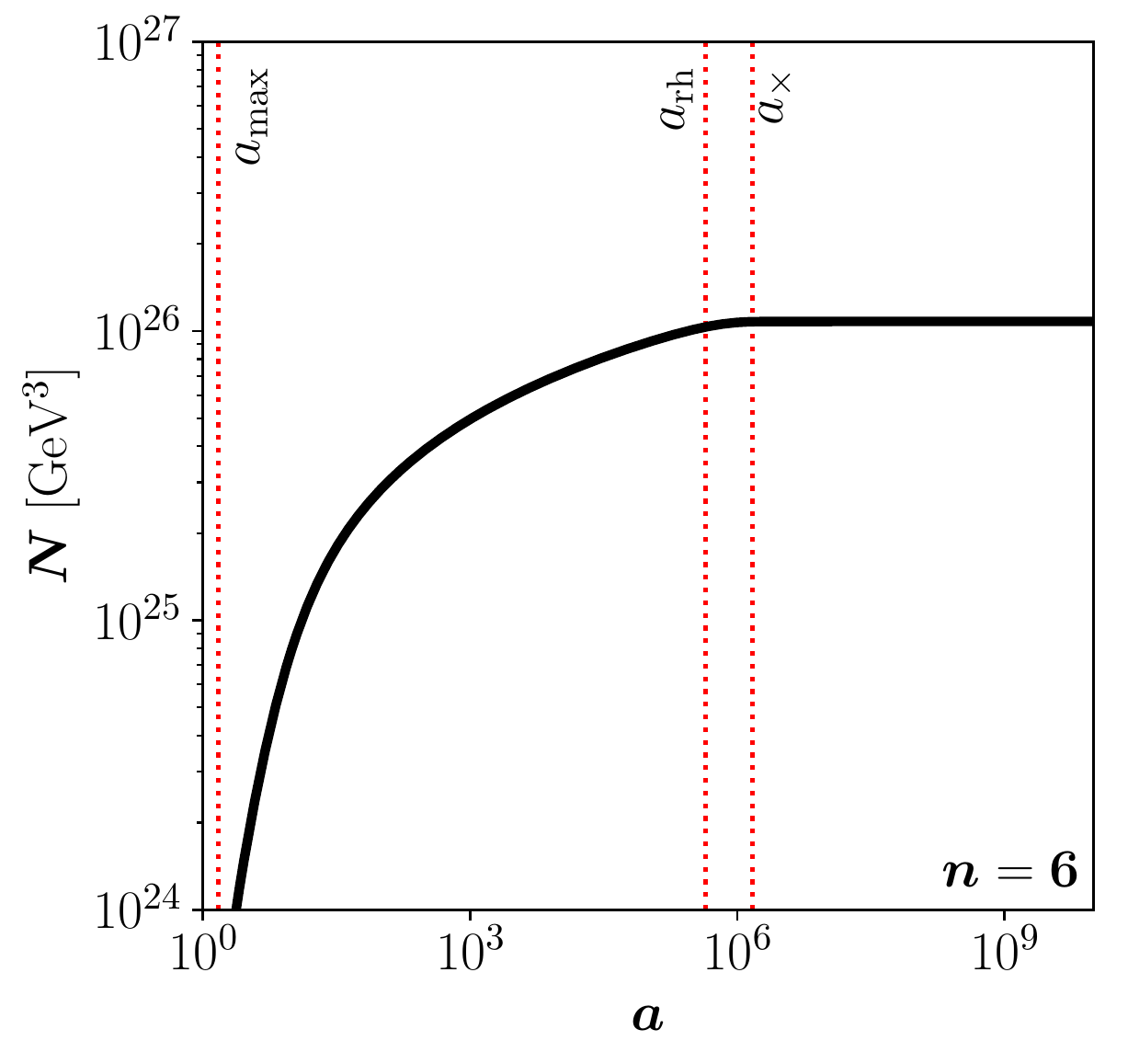}
\includegraphics[height=0.34\textwidth]{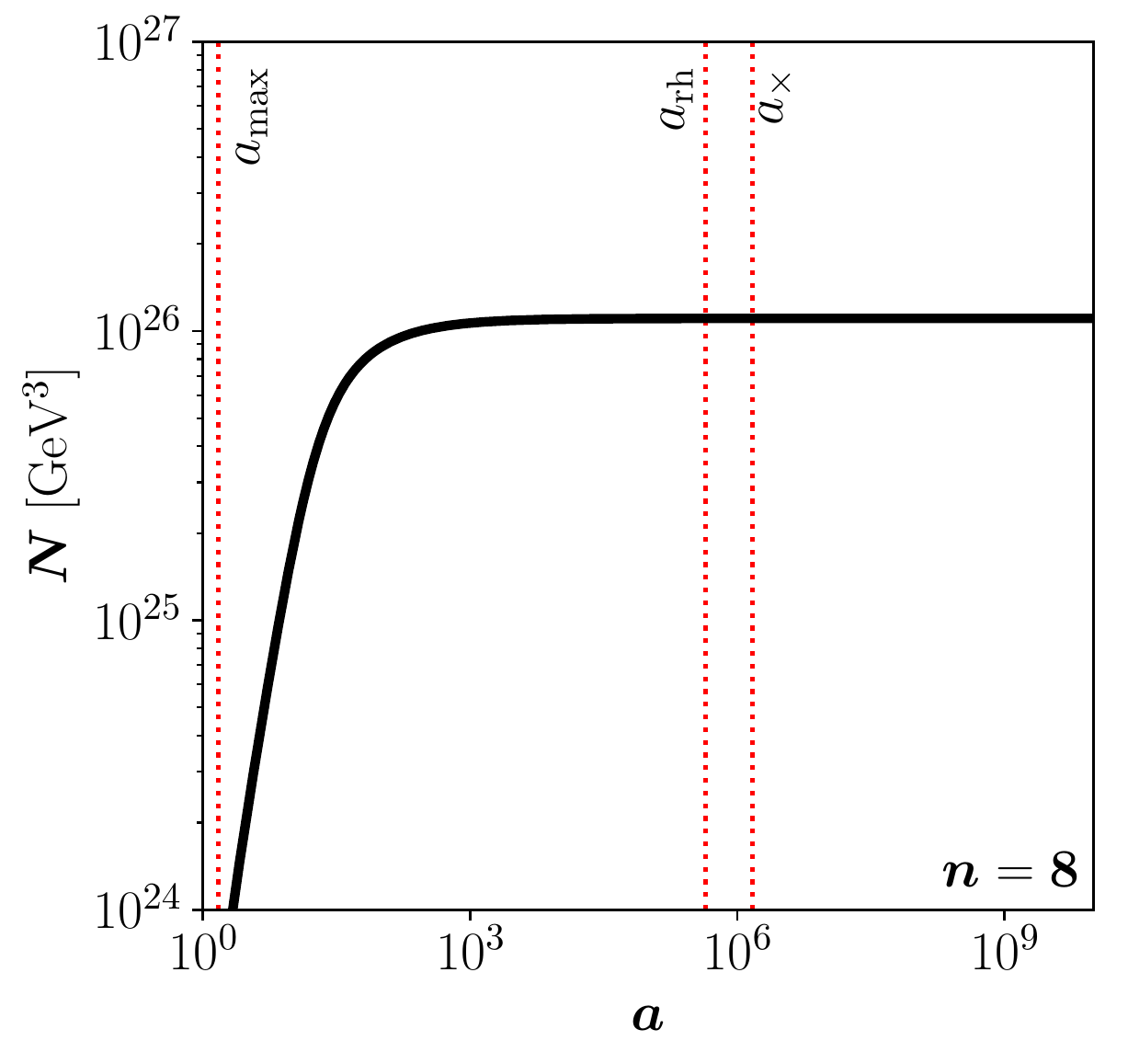}}
\centerline{\includegraphics[height=0.34\textwidth]{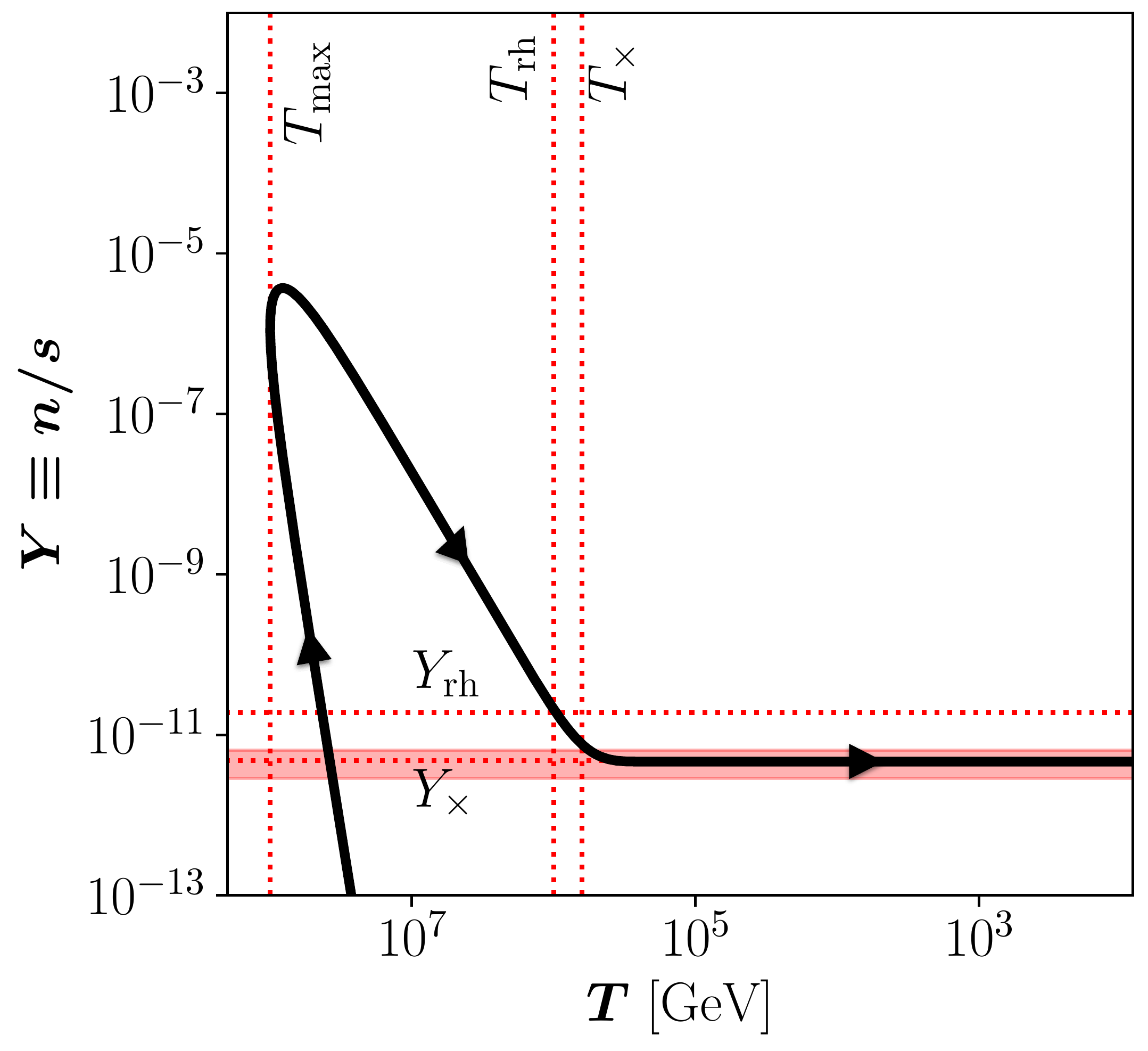}
\includegraphics[height=0.34\textwidth]{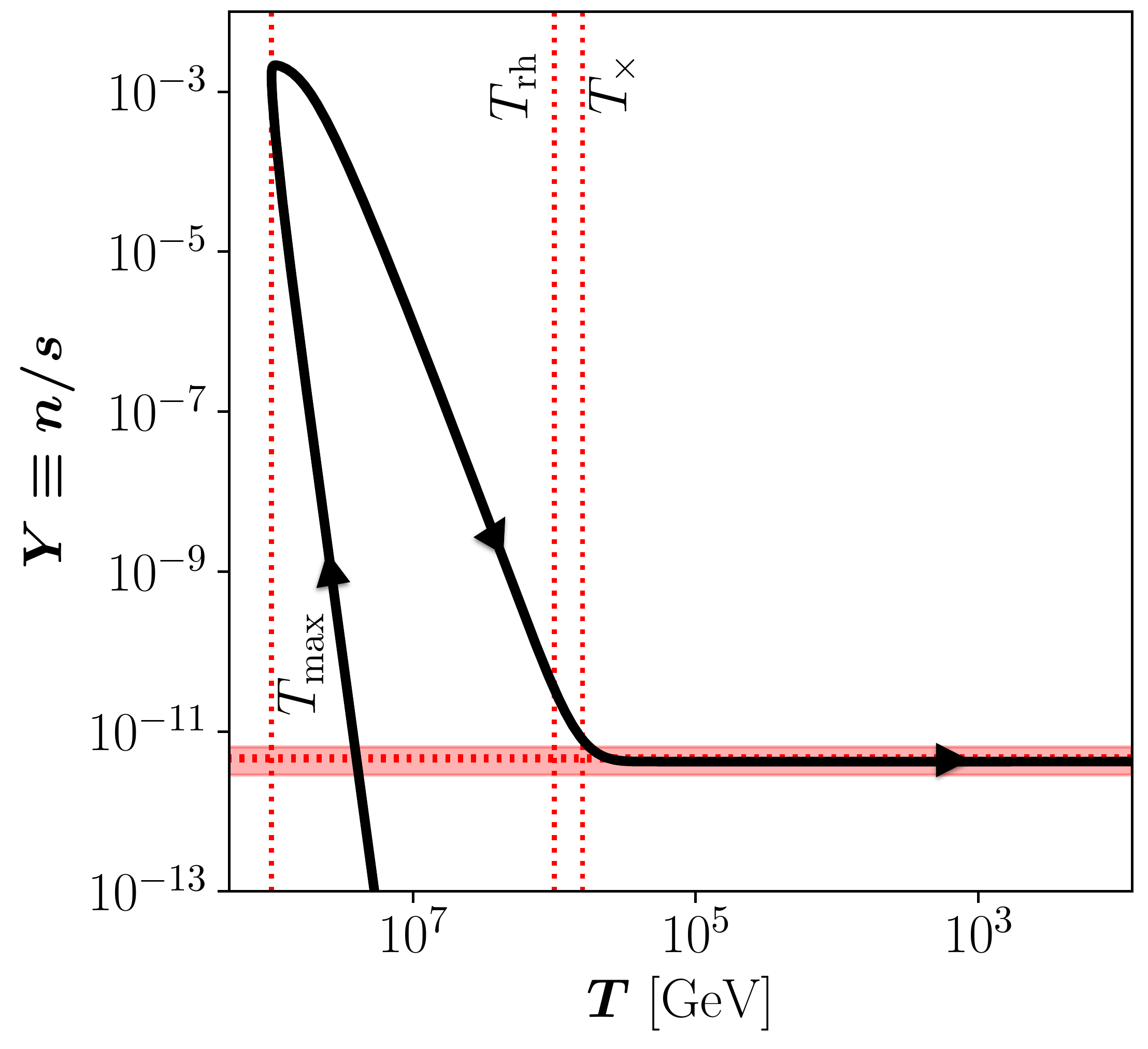}
\includegraphics[height=0.34\textwidth]{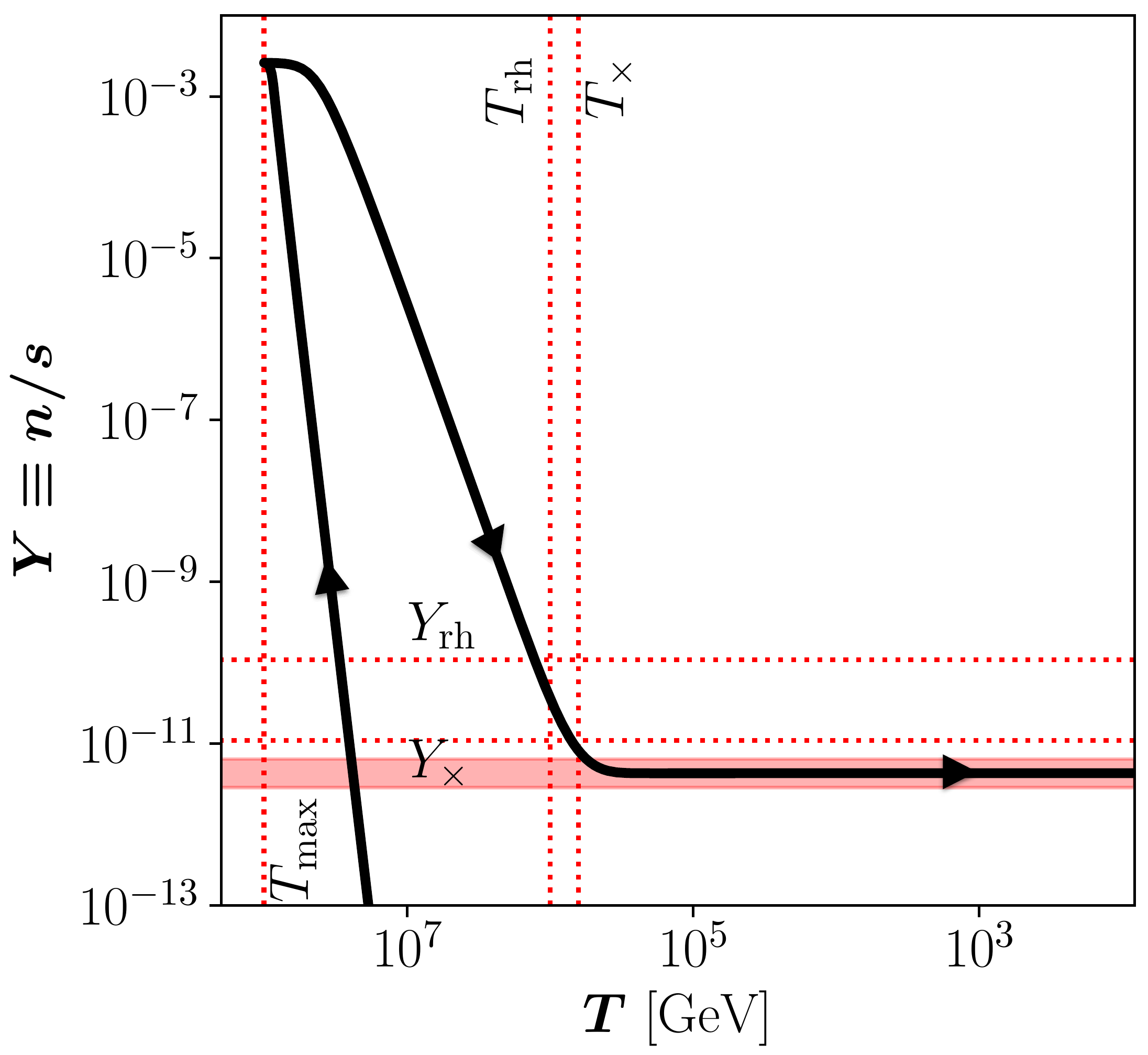}}
	\caption{Evolution of the DM comoving number density $N$ as a function of the scale factor (upper panels) and the DM yield $Y$ as a function of $T$ (lower panels), for values $\omega=0$, $\Trh=10^6$~GeV, $\Tmax=10^8$~GeV and $m=100$~GeV.
	The left panels correspond to $n=4$ and $\Lambda=1.4\times 10^9$~GeV, the central panels to $n=6$ and $\Lambda=2.8\times 10^8$~GeV and the right ones to $n=8$ and $\Lambda=1.6\times 10^8$~GeV. The values for $\Lambda$ were chosen in order to fit the observed DM abundance with the red bands showing the observed DM relic abundance today. The horizontal dotted lines depict the approximate numerical solutions $Y_\text{rh}$ and $Y_\times$. The dotted vertical lines correspond to $a=\amax$, $\arh$, $\ac$ and $T=\Tmax$, $\Trh$, $\Tc$ respectively. The arrows in the lower panels are directed to indicate the evolution with time.
}
	\label{fig:evolutionphi}
\end{figure}

In Figure~\ref{fig:evolutionphi} (upper panels) we present a number of examples which illustrate the evolution of the DM comoving number density $N$, as a function of the scale factor $a$, during the transition from matter domination ($\omega=0$) to radiation domination. It can be seen that for $n<n_c=6$ (left panels) the bulk of the DM relic abundance is produced near $T=\Trh$. On the contrary, for $n>n_c$ (right panels) the maximal production occurs between $\Tmax$ and $\Trh$. We also show the change in the DM yield $Y$ with temperature $T$ (lower panels), for $\omega=0$, $\Trh=10^6$~GeV, $\Tmax=10^8$~GeV, and $m=100$~GeV. In each case we chose the parameter $\Lambda$ such that the observed DM relic density is reproduced at late time. The horizontal dotted lines depict the approximate numerical solutions $Y_\text{rh}$ and $Y_\times$; the full numerical solution is better fitted by the analytical solution for $T\to\Tc$ rather than $T\to\Trh$.

\begin{figure}[t!]
\centerline{
\includegraphics[height=0.3\textwidth]{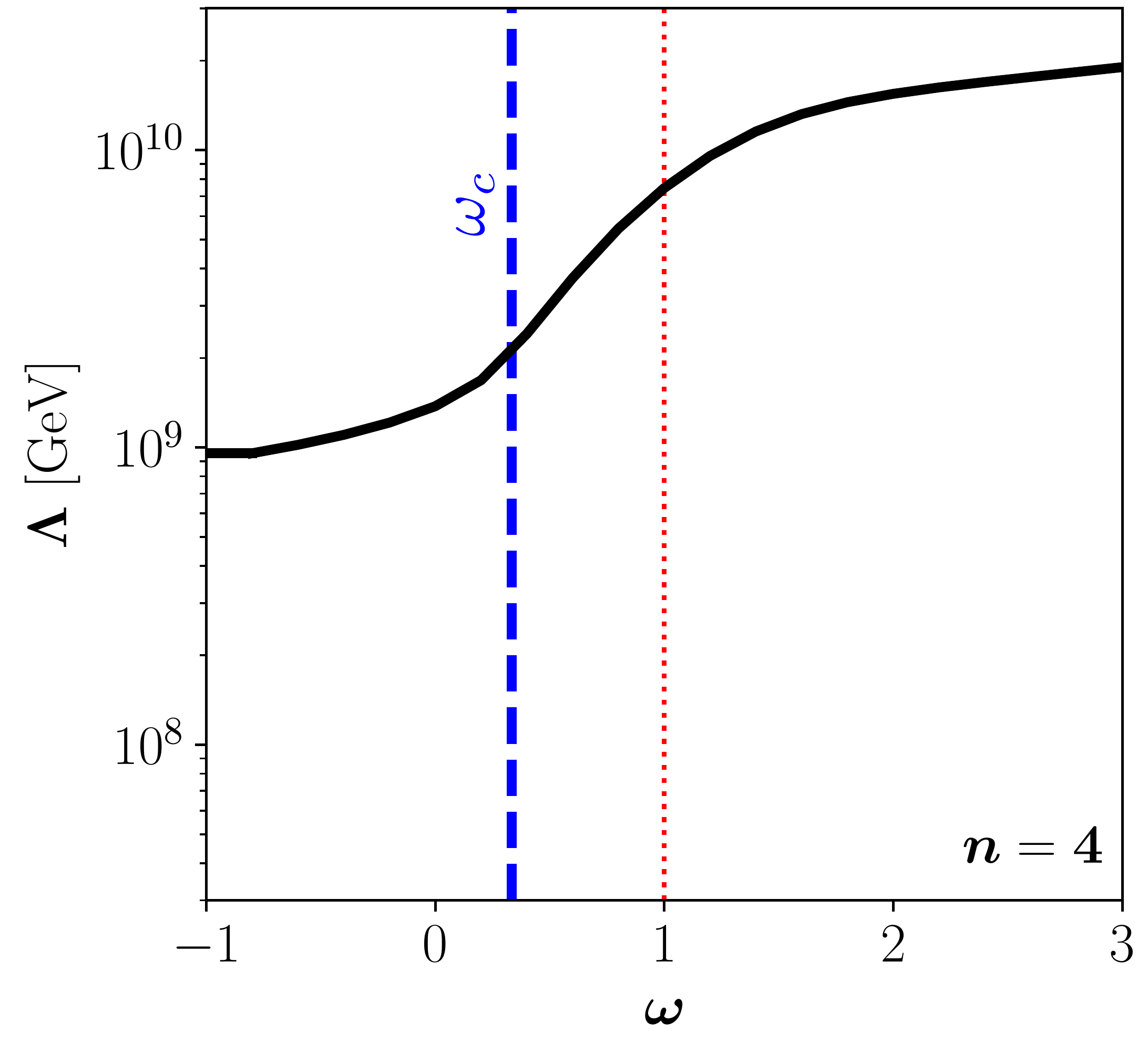}
\includegraphics[height=0.3\textwidth]{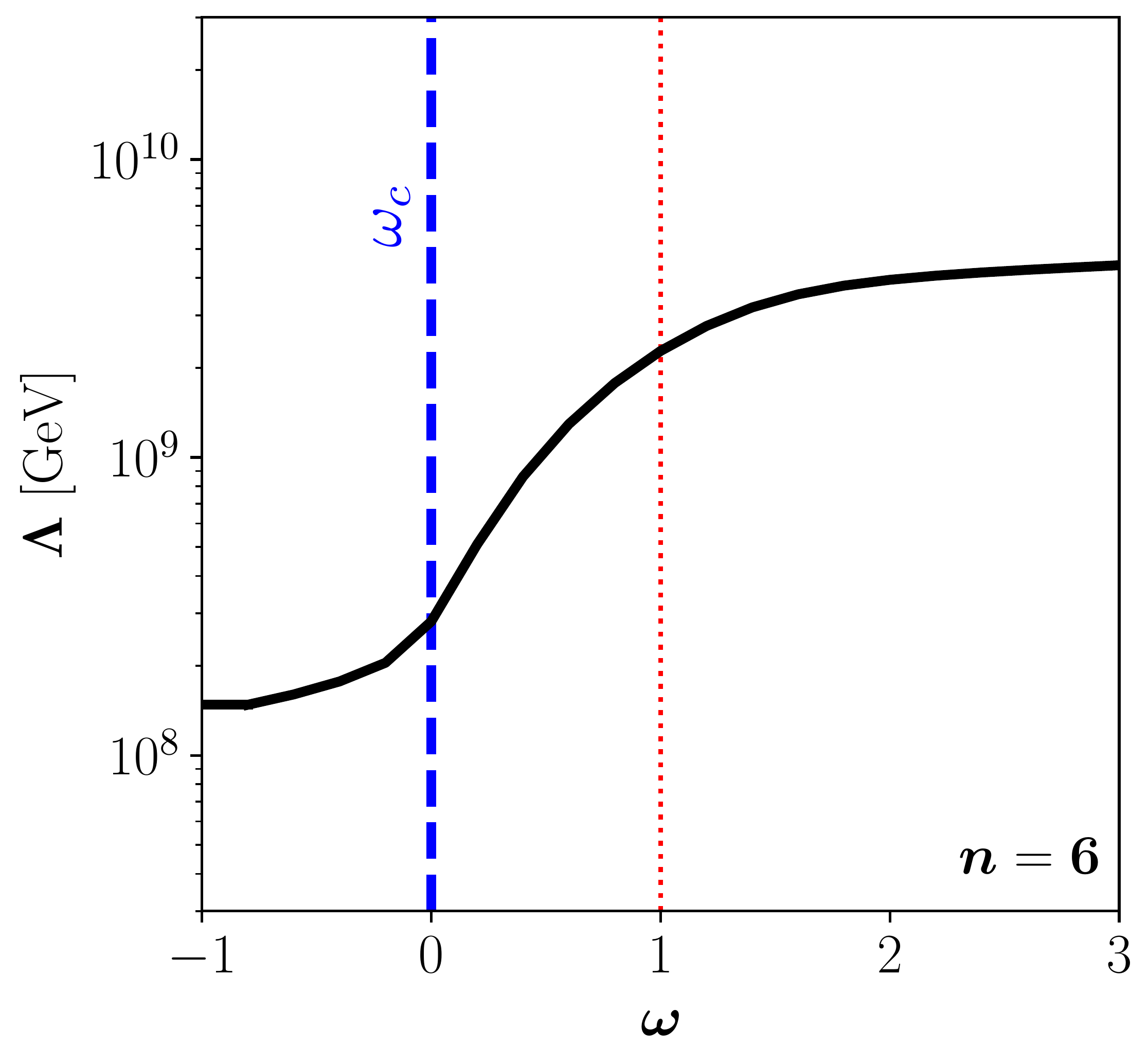}
\includegraphics[height=0.3\textwidth]{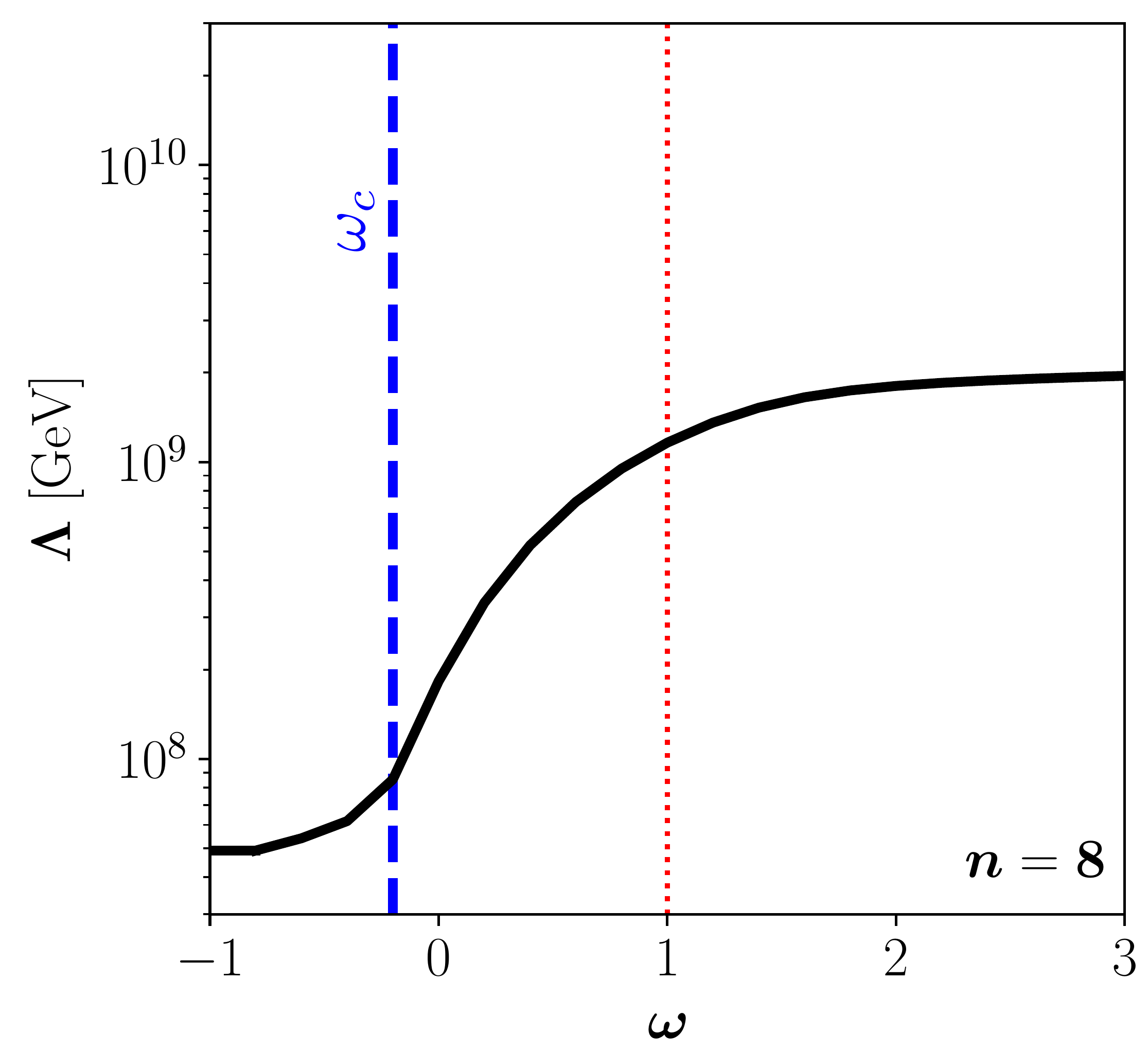}}
\vspace{-2mm}
	\caption{Contours in the $\Lambda-\omega$ space that generates the observed DM abundance for $m=100$~GeV, $\Trh=10^6$~GeV, $\Tmax=10^8$~GeV for $n=4$ (left), $n=6$ (central) and $n=8$ (right).
	The dashed blue lines correspond to $n=n_c$ (or equivalently to $\omega=\omega_c$), and the red dotted lines indicate $\omega=1$, since value $\omega>1$ are based on certain special classes of cosmological models, e.g.~\cite{DEramo:2017gpl,Gardner:2004in,Choi:1999xn,Dutta:2016htz,Okada:2004nc,Meehan:2014bya}.
\vspace{-1mm}	}
	\label{fig:La-om}
\end{figure}

Additionally, in Figure~\ref{fig:La-om}, we present  the parameter space that generates the observed DM abundance for DM of mass $m=100$~GeV, reheating temperature $\Trh=10^6$~GeV, and maximum temperature $\Tmax=10^8$~GeV for $n=4$, 6, 8.
These mass dimensions correspond to critical values of the equation of state $\omega_c=1/3$, 0 and -1/5,
where $\omega_c\equiv\frac{6-n}{2+n}$ comes from eq.~\eqref{eq:nc} and denotes the critical value of the equation of state $\omega_c$.

Observe that for $\omega<\omega_c$ (left of the blue lines) the DM relic abundance is produced after the decay of the inflaton and therefore the sudden decay approximation works well. That implies that the properties of the inflaton and in particular its equation of state have a marginal impact on the final DM density. In contrast, for $\omega>\omega_c$ DM is mainly generated between $\Tmax$ and $\Trh$, during the decay of the inflaton, and hence an important dependence on the equation of state is present. To maintain the observed DM abundance within the observed limits, the enhanced production during the decay of the inflaton for larger values of $\omega$ must be compensated by increasing the scale  $\Lambda$.


\subsection{Boost factors for dark matter production}
\label{ss:boost}

This work can be viewed as a generalization of the analysis of~\cite{Garcia:2017tuj} to $\omega$ different from zero. Thus to make contact with this earlier work we emulate their approach of characterising the impact on DM by defining a boost factor $B$ for the DM relic density which is the ratio of the DM abundance taking into account non-instantaneous reheating relative to the abundance in the instant decay approximation. Reevaluating eqs.~\eqref{eq:Yrh} and~\eqref{eq:Yrhcrit} at the temperature $T=\Trh$ (instead of $T=\Tc$), for a given equation of state, and comparing to eq.~\eqref{eq:Y0} implies an enhancement of the DM relic density in the non-instantaneous case given by 
\beq
	B\simeq\begin{cases}
		\frac13\frac{(1+n)(2+n_c)}{n_c-n} & \text{ for } n<n_c\,,\\[8pt]
		\frac{(1+n)\,(2+n)}{3}\ln\frac{\Tmax}{\Trh} & \text{ for } n=n_c\,,\\[8pt]
		\frac13\frac{(1+n)(2+n_c)}{n-n_c}\left[\frac{\Tmax}{\Trh}\right]^{n-n_c} & \text{ for } n>n_c\,.
	\end{cases}
	\label{nn}
\eeq
Conversely, one can express the boost factor as a condition on $\omega$ for a given $n$ as follows
\beq
	B\simeq\begin{cases}
		\frac13\frac{7-\omega_c}{\omega_c-\omega} & \text{ for } \omega<\omega_c\,,\\[8pt]
		\frac83\frac{7-\omega}{(1+\omega)^2}\ln\frac{\Tmax}{\Trh} & \text{ for } \omega=\omega_c\,,\\[8pt]
		\frac13\frac{7-\omega_c}{\omega-\omega_c}\left[\frac{\Tmax}{\Trh}\right]^\frac{8(\omega-\omega_c)}{(1+\omega)(1+\omega_c)} & \text{ for } \omega>\omega_c\,,
	\end{cases}
\eeq
where $\omega_c\equiv\frac{6-n}{2+n}$ denotes the critical value of the equation of state such that for $\omega>\omega_c$ DM produced by via an operator of mass dimension $5+n/2$ will be parametrically enhanced. 

\begin{figure}[t!]
\begin{center}
\includegraphics[height=0.35\textwidth]{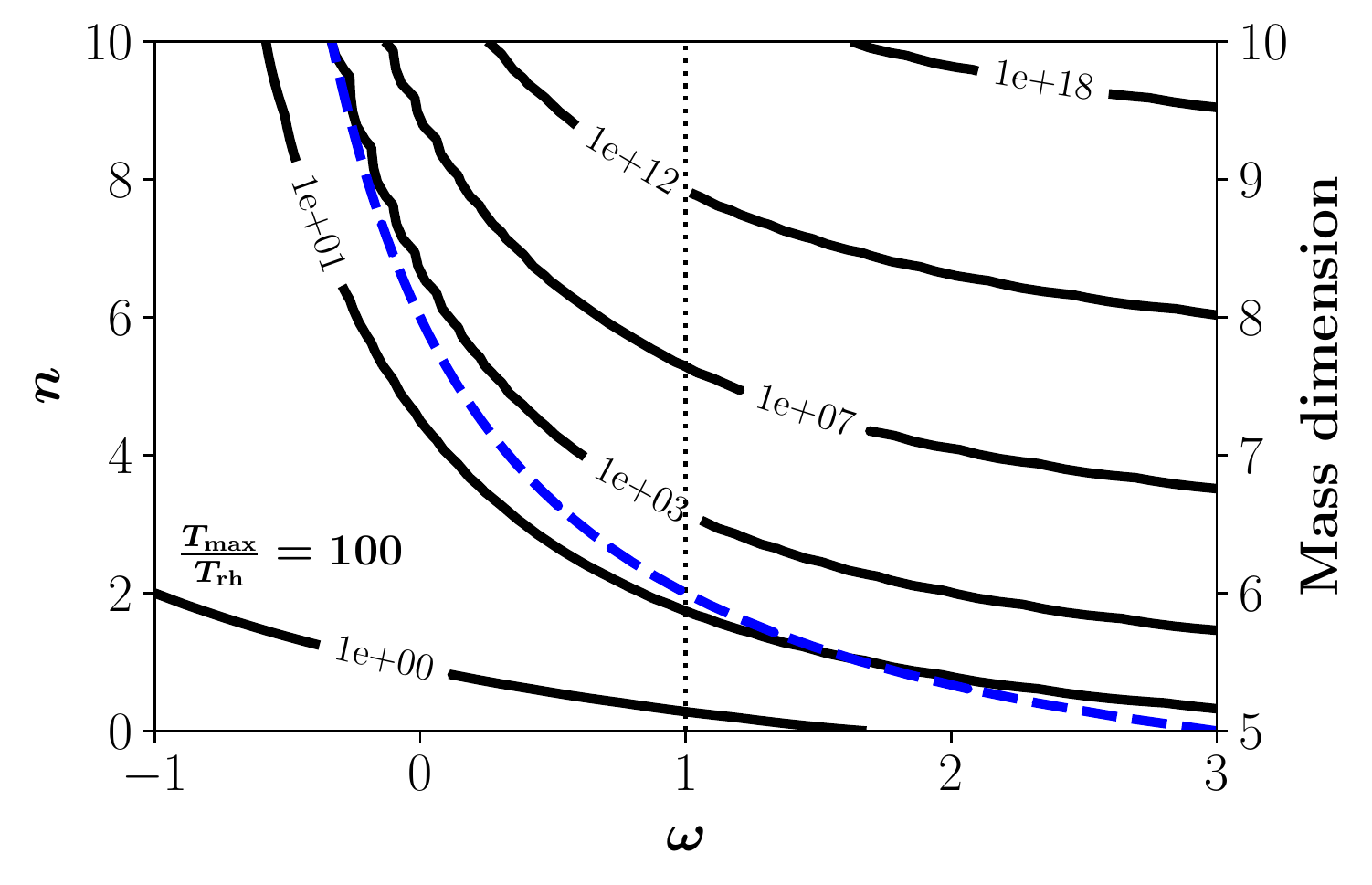}
	\vspace{-5mm}	
	\caption{Taking $\Tmax/\Trh=100$ we show a contour plot of the boost factor $B$  in the $\omega-n$ plane, where $n$ corresponds to the temperature dependence of the  cross section $\sv\sim T^n/\Lambda^{2+n}$. Equivalently the exponent value $n$ corresponds to UV freeze-in via an effective operator of mass dimension $5+n/2$.  For ease of conversion, the right hand axis gives the corresponding mass dimension of the portal operator for each value of $n$.
 The boost factor $B$ characterizes the enhancement to the relic density due to reheating effects by normalizing to the expectations from instantaneous reheating as discussed in Section \ref{ss:boost}.  The dashed blue lines correspond to the critical threshold $n=n_c$, beyond which (for $n>n_c$) the DM relic density due to UV freeze-in via an operator of mass dimension $n>n_c$ is parametrically enhanced. The vertical line indicates $\omega=1$.
\vspace{-5mm}	}
	\label{fig:boost0}
\end{center}
\end{figure}

\begin{figure}[t!]
\centerline{
\includegraphics[height=0.32\textwidth]{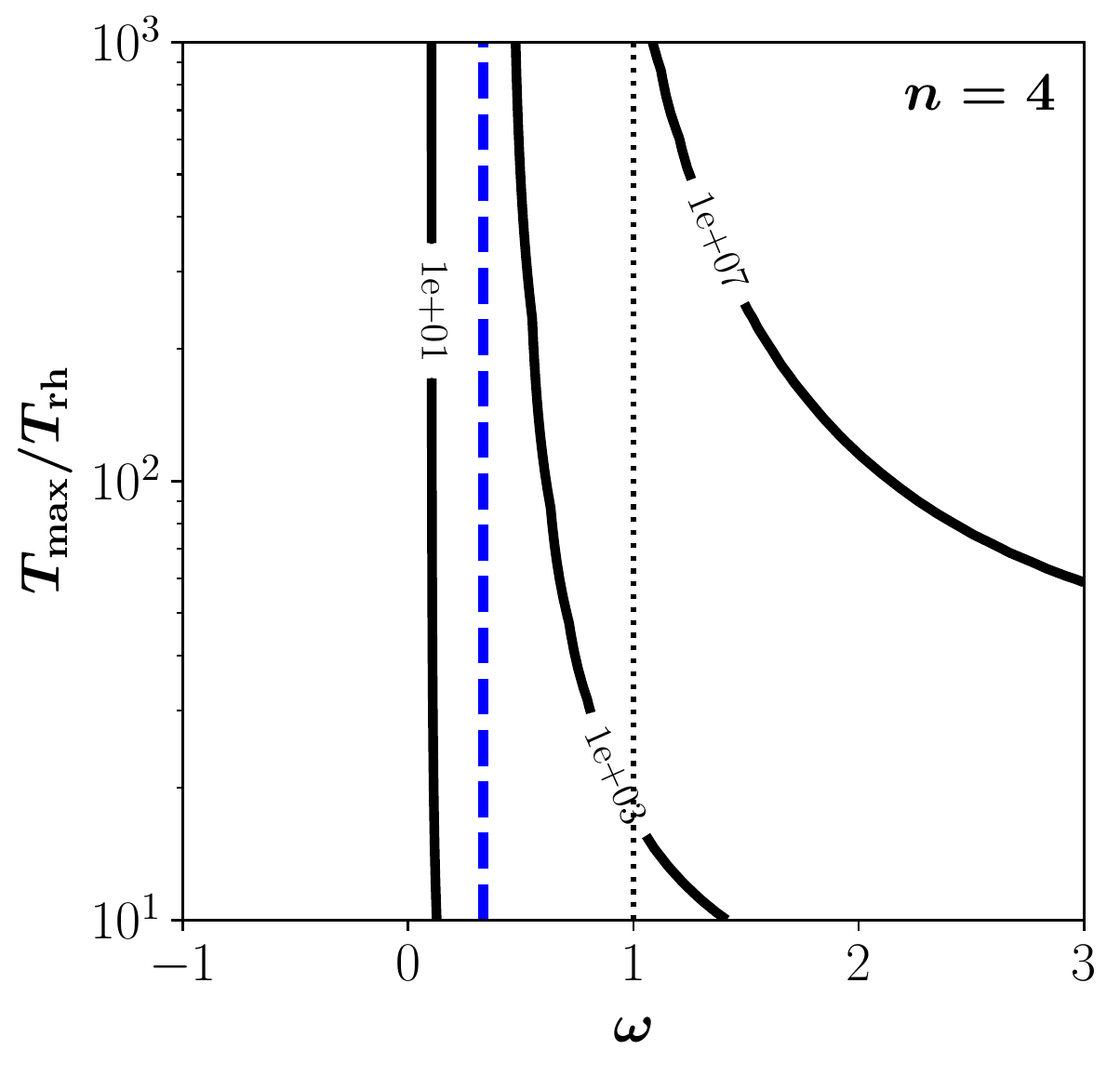}
\includegraphics[height=0.32\textwidth]{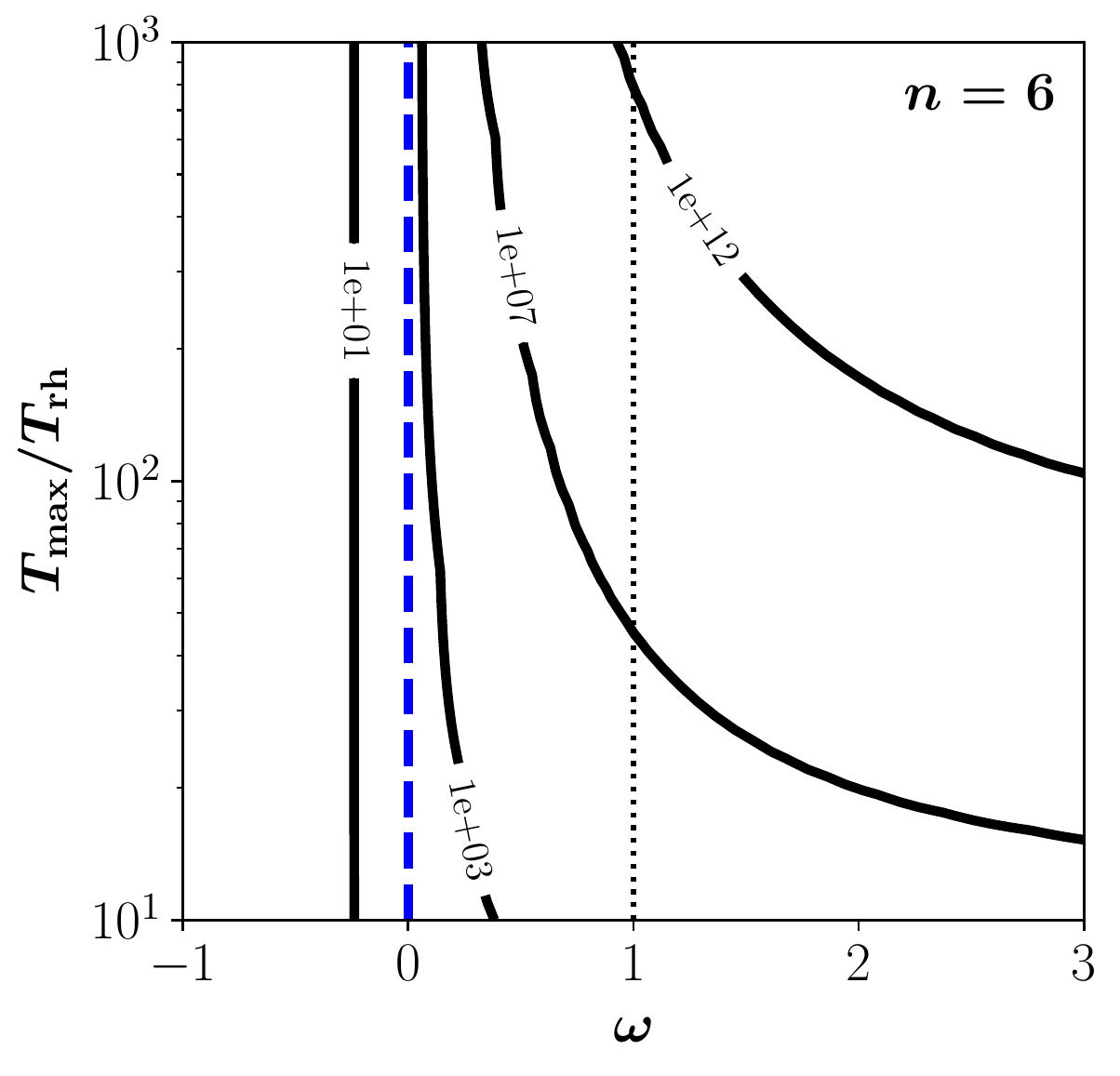}
\includegraphics[height=0.32\textwidth]{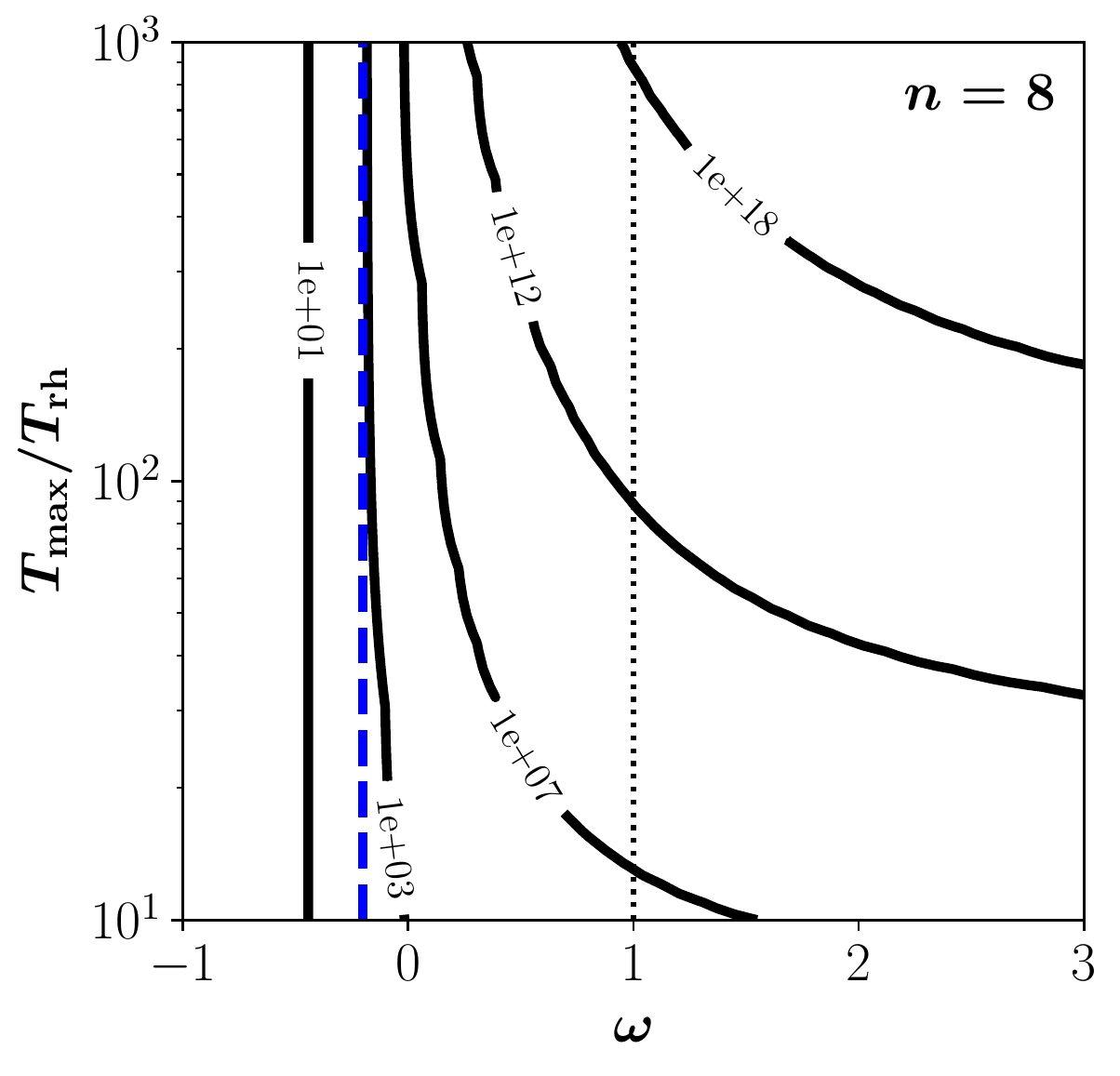}}
\vspace{-3mm}
	\caption{Contours of the boost factor $B$ in the $\omega-\Tmax/\Trh$ plane, where $\omega$ is the equation of state prior to reheating. We present plots for three different choices of the cross section temperature dependence $\sv\sim T^n/\Lambda^{2+n}$, for $n=4$, 6, 8. 
	The dashed vertical line corresponds to $\omega=1$.
	}
	\label{fig:boost}
\vspace{-3mm}
\end{figure}

This  boost factor is a clean way to characterise the enhancement since many of the other factors fall out due to the similarities of the underlying particle physics model. In particular, we highlight that the boost factors only depend on $n$, $\omega$ and the ratio $\Tmax/\Trh$, but not on $\Lambda$ or $m$.  The case $\omega=0$ in eq.~(\ref{nn}) agrees with Garcia-Mambrini-Olive-Peloso~\cite{Garcia:2017tuj} up to a factor of $5/3$ which arises due to differing definitions of the exact point of reheating.
We also check that the DM abundance dominantly arises via the lowest dimension freeze-in operator present, and, as expected from an effective field theory perspective, successive higher dimension operators give smaller contributions even accounting for the boost factor.

In Figure~\ref{fig:boost0} we show contours for the boost factor $B$ in the $\omega-n$ plane, taking a fixed value for $\Tmax/\Trh=100$ and in Figure~\ref{fig:boost} we show $B$ contours in the $\omega-(\Tmax/\Trh)$ plane, for $n=4,6,8$. Interestingly the boost factors can easily be several orders of magnitude, however the adjustment to $\Lambda$ needed to match the DM relic density taking into account the boost is much smaller (cf.~Figure \ref{fig:La-om})
 due to the strong dependence $N\propto\Lambda^{-(n+2)}$ in eq.~\eqref{eq:N1}.

\subsection{Loop induced production of dark matter}

Thus far we have assumed that direct and loop induced production of DM via decays of $\phi$ can always be neglected, in this subsection we will quantify when this assumption is reasonable, drawing on the study of~\cite{Kaneta:2019zgw} for the case of an early matter dominated period. While we might reasonably suppose that a direct coupling to the inflation $\phi$ can be effectively absent due to very small couplings, or some symmetry or special construction which forbids the operator, the loop induced contribution involving the Standard Model states is unavoidable. Thus we should quantify when loop induced DM production due to $\phi$ decays is important and identify in which regions of parameter space such contributions can be safely neglected.

Including the radiative decay of the inflaton to DM particles means that we have to modify eq.~(\ref{eq:cosmo1}) to 
\beq\label{eq:philoop}
\frac{dn}{dt}+3\,H\,n=-\sv\left(n^2-n_\text{eq}^2\right) +\frac{\rho_\phi}{m_\phi} \Gamma_\phi \text{Br},
\eeq
where Br represents the branching ratio of the inflaton to DM particles, and $m_\phi$ is the mass of   the decaying state $\phi$. We assume here that the $\phi$ number density is given by $n_\phi=\rho_\phi/m_\phi$.

To ascertain when the population of DM due to decays is non-negligible relative to that due to UV freeze-in we calculate the contribution to the yield due to decays $Y_D$ and compare it to the UV freeze-in yield of eq.~(\ref{eq:Yrh}) which we shall label $Y_{\rm FI}$ below.
To calculate the contribution to the yield from loop induced decays we neglect the first term on the RHS of eq.~\eqref{eq:philoop} and following an analogous procedure as the one in Section~\ref{general-nst}, we can express the evolution of the comoving DM number density due to $\phi$ decays $N_D(a)$  as
\begin{equation}
	\frac{dN_D}{da}=\frac{\rp\,\Gp\,a^2}{\mphi\,H}\br~.
\end{equation}
For $\omega\ne 1$ this admits the following analytical solution
\beq
N_D(T) &=\frac{2\,\br}{1-\omega}\frac{\Gp\,\Mp^2\,\Hini}{\mphi}\,\aini^3 \left[\left( \frac{\Tmax}{T}\right)^{4\frac{1-\omega}{1+\omega}}\left[ \frac{8}{3(1+\omega)}\right]^{3\frac{1- \omega}{5-3 \omega}} -1 \right].
\eeq
Using eqs.~\eqref{amax}, \eqref{Hini}, and~\eqref{eq:convert} one can convert $N_D$ to the contribution to the yield due to loop induced decays, and evaluating this
at $T=\Trh$ we obtain for $\omega\neq1$ the following\footnote{For the case $\omega=1$ one finds $Y_D(\Trh)=3\frac{\gs}{\gss}\frac{\Trh}{\mphi}\,\br\,\ln\left[\frac43\left(\frac{\Tmax}{\Trh}\right)^{4/3}\right]$.}
\begin{equation}
	Y_D(\Trh)=2\frac{\gs}{\gss}\,\frac{\Trh}{\mphi}\br \left[ \frac{1}{1-\omega}- \frac{1}{1-\omega}\left(\frac{\Trh}{\Tmax}\right)^{4\frac{1-\omega}{1+\omega}}\left[\frac{3(1+\omega)}{8}\right]^{3\frac{1-\omega}{5-3\omega}} \right].
	\label{YD}
\end{equation}
For $\Trh/\Tmax<0.1$ the term in the brackets is $\mathcal{O}(1)$ for $-1<\omega\lesssim0.5$ for larger values of $\omega$ this term can grow large. This can be more clearly seen by defining a boost factor $B_D$ similar to previously to compare the contribution to the yield from decays in the non-instantaneous case, given by eq.~\eqref{YD}, and the contribution found using the instantaneous approximation.

For sudden decays of $\phi$ the energy density of $\phi$ at $T=\Trh$ is shared between the Standard Model bath and the DM according to the relative branching fractions. Since we assume that prior to $\phi$ decays the energy density of DM and radiation are negligible it follows that $\rho_{\rm DM}(\Trh)=\rho_\phi(\Trh)\,\br$ and $\rR(\Trh)=\rho_\phi(\Trh)(1-\br)\simeq \rho_\phi(\Trh)$, since we defined $\br$ to denote the  $\phi$  branching ratio to DM and as the decays to DM are loop induced $\br\ll1$. If we further suppose that each  $\phi$ decay produces two DM particles then it follows that $n_D(\Trh)=2\,\br\,\rp(\Trh)/\mphi$ and the yield is therefore
\begin{equation}
	Y_D(\Trh)\bigg|_{\rm sudden}\simeq2\frac{\rR(\Trh)}{\mphi\,s(\Trh)}\br=\frac34\frac{\gs}{\gss}\frac{\Trh}{\mphi}\br.
\label{YDS}
\end{equation}
Then the boost factor is the ratio of eq.~\eqref{YD} and eq.~\eqref{YDS} given by (for $\omega\neq1$)
\beq
B_D= \frac{8}{3(1-\omega)} \left(1- \left(\frac{\Trh}{\Tmax}\right)^{4\frac{1-\omega}{1+\omega}}\left[\frac{3(1+\omega)}{8}\right]
 ^{3\frac{1-\omega}{5-3\omega}} \right),
\eeq
and we observe that for $\Trh/\Tmax<0.1$ the boost factor is $\mathcal{O}(1)$ for $-1<\omega\lesssim0.5$ but can be significant for $\omega>1$ and grows with the ratio of $\Tmax/\Trh$.

We now return to the comparison of the freeze-in yield and the contribution from loop induced decays. In particular, we will highlight that the reheat temperature is tied to the energy density of the inflaton, which in turn factors into the DM abundance due to decays. 
The branching ratio of the inflaton to DM depends on the model, but we can consider a simple model to obtain an intuition about the characteristic requirements. Let us consider a toy model in which a state $\psi$ is the proxy for the particles in the Standard Model bath, such that the inflaton coupling to the bath and UV freeze-in portal involving a fermion DM state $\chi$ can be described by the Lagrangian
\beq
\mathcal{L} = y_\psi  \phi \bar \psi \psi + \frac{1}{\Lambda^2} (\bar \psi \psi)( \bar \chi \chi).
\label{lag}\eeq 
This implies a radiative decay of $\phi$ to $\chi\overline{\chi}$ via a $\psi$ loop with a decay width (for $m_\psi < m_\phi$)
\beq 
\Gamma_{\phi \to \bar \chi \chi} \simeq \frac{y_\psi^2}{(4 \pi)^3} \frac{  m_\phi^5}{\Lambda^4}.
\eeq
The decay width of the inflaton to $\psi$ pairs is much larger and is approximately the total decay width of the inflaton
$\Gamma_\phi \simeq y_\psi^2 m_\phi/(8 \pi)$ and thus the $\phi$ branching ratio to DM is
\beq
\text{Br}  \simeq \frac{ m_\phi^4}{16 \pi^2 \Lambda^4}.
\eeq
Moreover, the reheat temperature of the thermal bath is given by $T_{\rm RH}\sim \sqrt{\Gamma_\phi M_{\rm Pl}}$ and thus
\beq 
\Trh^2 = \frac38\frac{y_\psi^2}{\pi^2} \sqrt{\frac{10}{\gs}} \Mp \,m_\phi.
\eeq

The requirement that  loop induced DM production is subdominant to the UV freeze-in contribution to the DM relic density implies $Y_D\ll Y_{\rm FI}$ and this places a restriction on the size of the branching ratio. We compare eqs.~(\ref{eq:Yrh}) and (\ref{YD}) with $n= 2$, and using the above we express the  branching ratio restriction in terms of the Lagrangian parameter of eq.~(\ref{lag}). For $\omega\neq1$ this condition can be written as a requirement on the $\phi$ mass 
\beq
m_\phi \ll \frac{g\,y_\psi}{g_\star } \Mp\times  
\begin{cases}
1 &  ~~{\rm for}~ -1<\omega<1, \\
\left[\frac{3}{8}(1+\omega)\right]^{3\left(\frac{1-\omega}{3\omega-5}\right)} & ~~ {\rm for}~\omega>1.
\end{cases}
\eeq
Therefore, in this simple model, provided $ y_\psi \sim\mathcal{O}(1)$, $\omega\sim1$ and $m_\phi < \Mp$ typically the DM  abundance due to the decay of the inflaton can be safely neglected and DM production via UV freeze-in sets the DM relic density.


\section{Models}
\label{S3}

In this section we explore various implementation in which the initial equation of state can be important for UV freeze-in. Specifically, we focus on 
\begin{itemize}
\item[\S\ref{3.1}.] Gravitino production,
\item[\S\ref{3.2}.] Spin 2 portal,
\item[\S\ref{3.3}.] Moduli portal dark matter,
\item[\S\ref{3.4}.] Higgs portals.
\end{itemize}

\subsection{Gravitino production}
\label{3.1}

In supersymmetric (SUSY) extensions of the Standard Model, the superpartner of the graviton is the gravitino. Notably, the gravitino is commonly the Lightest Supersymmetric Particle (LSP) and in R-parity conserving theories it is stable and thus a viable DM candidate. 
In local SUSY, the goldstino becomes the longitudinal component of gravitino and for very light gravitinos (much lighter than goldstino), the coupling to this longitudinal component dominates, and determines the evolution of gravitino DM in the early universe. In this limit the production of gravitinos from the thermal bath is through non-renormalizable operators suppressed by the SUSY breaking $F$ term with $\sv \propto T^6/F^4$.

Furthermore, in models of High Scale SUSY \cite{Hall:2009nd} the superpartners have masses well above the electroweak scale. The suppression of the gravitino mass $m_{3/2}$ to the mass scale $\widetilde m$ of the other superpartners is a natural consequence of gauge mediated SUSY breaking \cite{Giudice:1998bp}, but can also occur in certain gravity mediation constructions \cite{Benakli:2017whb}, and other mediation mechanisms of SUSY breaking. Thus one can readily envisage scenarios in which the gravitino is the only sparticle lighter than the inflationary reheating scale. 

Since the splitting between gravitino and superpartners can be relatively large the hierarchy $m_{3/2}\ll \Trh \ll \widetilde{m}$  is quite conceive and thus gravitinos. This hierarchy implies that the production of gravitinos from R-parity violating decays of superpartners will be negligible and the thermal production of gravitino pairs $\tilde G$ via freeze-in  sets the relic abundance. Hence, we consider $ X + Y \to \tilde G + \tilde G$, where $X$ and $Y$ are states in the thermal bath, with the production cross section  \cite{Garcia:2017tuj}
\beq
\sv \simeq 100 \times \frac{ T^6}{\Mp^4  m_{3/2}^4},
\label{grav1}
\eeq
where in the expression of the RHS we have adopted the gravity mediated scenario of \cite{Garcia:2017tuj,Benakli:2017whb} and identified $F\sim\Mp  m_{3/2}$

The authors of~\cite{Garcia:2017tuj} studied the case of gravitino production in the early universe during the reheating due to a transition from matter to radiation domination, and highlighted that in which case the assumption of instantaneous decay of the inflaton breaks down. Building on the model independent study of Section \ref{general-nst} we now look to generalise this to the case that the equation of states of the early universe is some arbitrary $\omega$.

Similar to the steps to the derivation in Section~\ref{general-nst} the gravitino comoving number density $N_{3/2}$ is found by solving eq.~\eqref{eq:cosmo1a} to obtain 
\beq
N_{3/2}(T) \simeq 
\frac{8  \times 10^2 \,g^2}{3 \pi^4\,(6-n_c)(1+\omega)}\frac{\Tc^{4\frac{3+\omega}{1+\omega}}}{\Mp^{4}m_{3/2}^4\Hini }\left[\frac{\ac^{3+\omega}}{\aini^{1+\omega}}\right]^\frac32 \left[\Tmax^{6-n_c}-T^{6-n_c}\right].
\eeq
The yield in the limit $ T \to \Tc$ is
\beq 
Y_{3/2}(\Tc) \simeq \frac{45 \times 10^{2} \, g^2}{ 2\pi^2 \gss \Mp^3 m_{3/2}^4}\sqrt{ \frac{10}{\gs}}
 \begin{cases}
		\frac{1}{\omega}\, \Tc^{\frac{7- \omega}{1+ \omega}}\left(\Tmax^{\frac{8\omega}{1+\omega}} - \Tc ^{\frac{8\omega}{1+\omega}}\right)  & \text{ for } \omega \neq 0,\\[12pt]
		\Tc^7 \ln \left(\frac{\Tmax}{\Tc}\right)& \text{ for } \omega= 0.
\end{cases}
\eeq 
To compare with the case of instantaneous inflaton decay, we can compute the boost factors:
\beq
B =
 \begin{cases}
\frac{7}{3 |\omega|} &\text{ for } \omega < 0,\\[12pt]
\frac{56}{3} \ln \left(\frac{\Tmax}{\Trh}\right)& \text{ for } \omega= 0,\\[12pt]
\frac{7}{3 \omega}\,  \left[\frac{\Tmax}{\Trh}\right]^{\frac{8 \omega}{1+ \omega}}  & \text{ for } \omega > 0.
\end{cases}  
\label{grav2}
\eeq
Moreover, it follows that the expected gravitino relic abundance for $\omega\neq0$ is given by  
\beq
\Omega_{3/2}
 \simeq 0.2 \times 
 \left( \frac{ \text{ 30 GeV}}{ m_{3/2}}\right)^3\left( \frac {\Tc}{10^7~\text{GeV}}\right)^7  
\frac{1}{\omega}  \left[ \left(\frac{\Tmax}{\Tc}\right)^{\frac{8\omega}{1+ \omega}} -1 \right]. \eeq

While for low scale SUSY gravitinos are typically produced in association with another superpartner, in an $M_{\rm Pl}$-suppressed process, this is not possible if the gravitino is the sole SUSY state below the reheating scale. If the other superpartners are too heavy to be produced then gravitinos must be paired produced which is a doubly suppressed process and as a result in this scenario gravitino DM  is generically underproduced. 
However, as highlighted in \cite{Garcia:2017tuj}, in the case that the inflaton decays and subsequent evolution is not well approximated as an instantaneous decay, then the DM relic abundance can be parametrically enhanced, potentially  adjusting the gravitino relic density to match the observed value. 

Observe from eq.~(\ref{grav2}) that enhancements of the gravitino DM abundance arise even for a non-instantaneous transition from matter ($\omega = 0$) to radiation domination in which case the abundance is logarithmically enhanced, and for non-standard cosmologies with $\omega > 0$, then its relic abundance can be greatly enhanced. For example, for an early period of kination domination (with $\omega = 1$) and assuming a ratio $ \Tmax/\Trh = 100$, then the relic abundance is enhanced by a factor of $B\sim10^8$. Indeed, for a concrete model $ \Tmax/\Trh$ may be bounded by the requirement that one does not overproduce gravitinos. 

\subsection{The spin-2 portal}
\label{3.2}

Freeze-in via a massless graviton was studied in~\cite{Garny:2017kha} and the scenario was subsequently extended in~\cite{Bernal:2018qlk} to the case of DM freeze-in via a massive spin-2 field. The typical way to couple a spin-2 field $\tilde{h}$ to matter is similar to the graviton, involving the energy momentum tensor $T_{\mu\nu}$ for the Standard Model and DM, via Lagrangian terms of the form  \cite{ArkaniHamed:2002sp, Katz:2005ir, Bernal:2018qlk}
\beq
\mathcal{L}\supset \frac{1}{\Lambda} \tilde{h}_{\mu\nu}(\lambda_{\rm SM} T^{\mu\nu}_{\rm SM}+\lambda_{\rm DM} T^{\mu\nu}_{\rm DM})~.
\eeq
For a graviton $\lambda_{\rm SM}=\lambda_{\rm DM}=1$, with $\Lambda=M_{\rm Pl}$ for a standard massless graviton, but where the scale $\Lambda$ can vary for a massive graviton. For a spin-2 field not related to gravity the couplings can in principle differ from unity, but here we restrict our considerations to the case $\lambda_{\rm SM}=\lambda_{\rm DM}=1$.

Such a heavy massive spin-2 mediator can potentially be identified with a massive graviton. While consistent theories of massive gravity can be constructed~\cite{deRham:2010kj,Hassan:2011hr,Dvali:2000hr}, observations constrain the graviton mass to be extremely small $m_2\lesssim 10^{-30}$~eV~\cite{deRham:2016nuf} which is not suitable for our considerations. Alternatively, there are classes of consistent bimetric gravity models~\cite{Hassan:2011zd} with one massless graviton and one massive graviton. For the (lesser studied) scenario that $m_2$ is much greater than Hubble constant today ($m_2\gg H_0$) the mass is largely unconstrained~\cite{Babichev:2016hir, Aoki:2016zgp}. Thus such bimetric gravity models provide a motivation for heavy spin-2 mediators with a large range of masses, such as the GUT scale or Planck scale.

We will consider two distinct cases depending on whether the mass $m_2$ of the spin-2 state $\tilde{h}$ is below or above the maximum bath temperature. Both cases lead to non-renormalisable operators if $\tilde{h}$ couples involving the energy momentum tensor, however the dimensionality of the operators differs as we highlight below.
In the case where the spin-2 mediator is lighter than the reheating temperature and can be produced on shell then the cross section for DM production due to bath interactions mediated  by $\tilde{h}$ is parametrically \cite{Bernal:2018qlk}
\beq
	\sv=\alpha \frac{T^2}{\Lambda^4}~,
	\label{light}
\eeq
where $\alpha$ is a numerical prefactor dependent on the spin of the DM, with 
values $\alpha\sim 0.19$ (spin 0), $2.04$ (spin 1/2) or $2.49$ (spin 1).\footnote{Note that for a massless graviton the cross section is given by eq.~(\ref{light}) with $\alpha\sim 1.28\times 10^{-2}$ (spin 0), $1.32\times 10^{-1}$ (spin 1/2) or $1.55\times 10^{-1}$ (spin 1), together with $\Lambda=\Mp$~\cite{Bernal:2018qlk}. Note that despite the massive mediator, the production rate are enhanced over massless gravitons exchange by the fact that $\Lambda< \Mp$.}
While one could absorb the numerical prefactor into $\Lambda$, these values are normalised such that $\alpha$ is correct for $\Lambda=M_{\rm Pl}$.
Carrying through similar calculations as above,  for the case  $m_2\ll T_{\rm max}$ the DM yield, which we denote $Y_{\rm light}$,  is found to be given by (for $\omega\neq1$)
\beq
	Y_{\rm light}(\Tc)\simeq
		\frac{45\,g^2\alpha}{\pi^7\,\gss}\sqrt{\frac{10}{\gs}}\frac{1}{\omega-1}\frac{\Mp\,\Tc^{\frac{7-\omega}{1+\omega}}}{\Lambda^4}\left[\Tmax^\frac{4(\omega-1)}{\omega+1}-\Tc^\frac{4(\omega-1)}{\omega+1}\right]~.
\eeq
Thus the yield is parametrically enhanced for $\omega>1$ and the boost factor is of the form
\beq
	B_{\rm light}\big|_{\omega>1}\simeq
		\frac{2}{\omega-1}\left[\frac{\Tmax}{\Trh}\right]^\frac{4(\omega-1)}{\omega+1}. 
\eeq

Conversely, in the case where the spin-2 mediator is heavier than the reheating temperature, but still constitutes  the dominant production channel
\beq
	\sv=\beta\frac{T^6}{\Lambda^4\,m_2^4}
\eeq
with the prefactor $\beta$ taking the values 
$\beta\sim 735$ (spin 0), $7814$ (spin 1/2) or $9505$ (spin 1).
 Observe that for $m_2>T_{\rm max}$ the thermally averaged production cross section is similar in form to that of the gravitino (cf. eq~\eqref{grav1}). 
 It follows that  the DM yield for $\omega\neq0$ in this case is
\beq
	Y_{\rm heavy}(\Tc)\simeq
		\frac{45\,\beta\,g^2}{2\pi^7\,\gss}\sqrt{\frac{10}{\gs}}\frac{1}{\omega}\frac{\Mp\,\Tc^{\frac{7-\omega}{1+\omega}}}{\Lambda^4\,m_2^4}\left[\Tmax^\frac{8\omega}{1+\omega}-\Tc^\frac{8\omega}{1+\omega}\right]. 
	\label{last1}
\eeq

Moreover, since this operator is the same dimensionality of the gravitino DM case, the boost factor corresponding to eq.~(\ref{last1}) is identical to that given in eq.~\eqref{grav2}, i.e.~for $\omega>0$
\beq
\label{eq:boost_spin2_2}
	B_{\rm heavy}\big|_{\omega>0}\simeq \frac{7}{3\omega}\left[\frac{\Tmax}{\Trh}\right]^\frac{8\omega}{1+\omega}.
\eeq
Since the dimension of the freeze-in operator changes depending on whether the spin-2 state is accessible or integrated out, the critical $\omega$ above which the abundance is parametrically enhanced changes. 
In particular, observe that the boost has a strong $T_{\rm max}$ dependence for $\omega>0$ with $m_2\gg\Tmax$, but requires $\omega>1$ in the case that $m_2\ll\Tmax$.

  \subsection{The moduli portal}
  \label{3.3}
  In models with extra dimensions, such as supergravity or string theory, light scalar fields known as moduli are associated with the compact dimensions. In principle such moduli could provide a portal between the Standard Model and DM~\cite{Chowdhury:2018tzw}. These moduli can be written as $\mathcal{T}\equiv \varphi+i\,a$ in terms of two real scalars $\varphi$ and $a$, and the wave function $\mathcal{Z}_k$, where $k$ is a Standard Model field that couples to this moduli field, can be expanded as
  $\mathcal{Z}_k\equiv1+\frac{\alpha_k}{\Lambda}\varphi+i\frac{\beta_k}{\Lambda}a$,
where $\alpha_k$ and $\beta_k$ are real constants and $\Lambda$ corresponds to the compactification scale. At leading order in $\Lambda^{-1}$ the moduli field couples to the Standard Model fermions $f$ and gauge fields $G$ via Lagrangian terms of the form  \cite{Chowdhury:2018tzw}
\beq
\mathcal{L}\supset &\frac{\alpha_H}{\Lambda}\varphi\left|D_\mu H\right|^2
- \frac{\alpha_H}{\Lambda}\mu_0^2 \varphi|H|^2+\left[\frac{1}{2\Lambda}\varphi\overline{f}i\gamma^\mu (\alpha_V^f-\alpha_A^f\gamma_5)D_\mu f+{\rm H.c.}\right]
\\
&+\frac{1}{2\Lambda}\partial_\mu a \overline{f}\gamma^\mu (\beta_V^f-\beta_A^f\gamma_5) f
-\frac{1}{4}\frac{\alpha_G}{\Lambda}\varphi G_{\mu\nu}G^{\mu\nu}
+\frac{2\beta_G}{\Lambda}\partial_\mu a \epsilon^{\mu\nu\rho\sigma} G_{\nu}\partial_\rho G^{\sigma}~,
\eeq
where $\mu_0$ is the Standard Model Higgs parameter.
The coupling of the moduli to the DM depends on the spin of the DM. Considering first the case of a scalar DM  state $S$ implies the following Lagrangian term
\begin{equation}
\mathcal{L}_S=\frac{\alpha_S}{\Lambda}\varphi\left|\partial_\mu S\right|^2,
\end{equation}
where $\alpha_S$ is the coupling constant of the real part of the moduli field to the scalar DM.
Similar to the spin-2 portal scenario explored above, there are two distinct cases depending on whether the mass of the modulus component $m_\varphi$ exceeds the maximum temperature of the thermal bath or not, and in each case the production cross section is parametrically 
\begin{equation}
\sv=\frac{\pi^4}{g^2\Lambda^4}\times
\begin{cases}
		\delta T^2     & \qquad m_\varphi\ll T_{\rm max}\,,\\[5pt]
		\delta \frac{ T^6}{ m_\varphi^4} &\qquad m_\varphi\gg T_{\rm max}\,,
	\end{cases}
\end{equation}
 where $\delta\propto\alpha_S^2\,\alpha_{\rm SM}^2$  in terms of $\alpha_{\rm SM}^2\equiv 2\alpha_H^2+3\alpha_G^2$. The constant of proportionality for $\delta$ differs for $ m_\varphi\ll T_{\rm max}$ and  $m_\varphi\gg T_{\rm max}$, however the precise value will be unimportant for our purposes, and for further details see \cite{Chowdhury:2018tzw}.

In the limit $m_\varphi\ll T$ the yield, which we denote $Y_{\rm light}$, at $T=\Tc$ is given by (for $\omega\neq1$)
\begin{equation}
Y_{\rm light}(\Tc)\simeq\frac{45}{\gss \pi}\sqrt{\frac{10}{\gs}}\,
\frac{\delta M_{\rm Pl}}{\Lambda^4\left(\omega-1\right)}
\, \Tc^{\frac{7-\omega}{1+\omega}} \left[\Tmax^{\frac{4(\omega-1)}{1+\omega}}-\Tc^{\frac{4(\omega-1)}{1+\omega}}\right]~.
\end{equation}
Thus in the case of scalar DM  with $m_\varphi\ll T_{\rm max}$ the abundance receives a $T_{\rm max}$ dependent boost  relative to the instant reheating approximation for  $\omega>1$  given by
\beq
	B_{\text{light}}\simeq
		\frac{2}{\omega-1}\left[\frac{\Tmax}{\Trh}\right]^\frac{4(\omega-1)}{\omega+1}. 
\eeq
Similarly, for the limit $m_\varphi\gg T$, in this case we have for the yield (for $\omega\neq0$)
\begin{equation}
Y_{\rm heavy} (\Tc)\simeq\frac{45}{2\pi^3\gss}\sqrt{\frac{10}{\gs}}\frac{\delta M_{\rm Pl}}{\Lambda^4 m_\varphi^4 \omega}
\Tc^{\frac{7-\omega}{1+\omega}} \left[\Tmax^{\frac{8\omega}{1+\omega}}-\Tc^{\frac{8\omega}{1+\omega}}\right].
\end{equation}
The corresponding boost factor for $\omega>0$ is given by
\beq
	B_\text{heavy}\simeq
		\frac{7}{3\omega}\left[\frac{\Tmax}{\Trh}\right]^\frac{8\omega}{\omega+1}. 
\eeq
Similar expressions are found for the case of vector boson DM with the difference that these can also receive a contribution mediated by $a$.

The case of freeze-in of fermion DM $\chi$ mediation via a moduli field provides a more interesting second example, the relevant Lagrangian contribution for which is given by~\cite{Chowdhury:2018tzw} 
\begin{equation}
\mathcal{L}_\chi=\left(\frac{1}{2\Lambda}\varphi\bar{\chi}i\gamma^\mu\left(\alpha_V-\alpha_A\gamma_5\right)\partial_\mu\chi+h.c.\right)+\frac{1}{2\Lambda}\partial_\mu a \bar{\chi}\gamma^\mu\left(\beta_V-\beta_A\gamma_5\right)\chi.
\end{equation}
There are again two distinct cases depending on whether the mediators (now both $a$ and $\varphi$ states can mediate interactions) can be produced by interactions in the  thermal bath and the production cross section for the case of fermion DM with mass $m$ is parametrically
\begin{equation}
\sv_j=\frac{\pi^4m^2}{g^2\Lambda^4}\times
\begin{cases}
\delta_j & \qquad m_j\ll T_{\rm max}\,,\\
\delta_j \frac{ T^4}{m_j^4} &  \qquad  m_j\gg T_{\rm max}\,,
\end{cases}
\end{equation}
with the effective couplings $\delta_j$ (for $j = \varphi$, $a$) given by $\delta_{\varphi}\propto \alpha_V^2\,\alpha_\text{SM}^2$ and $\delta_{a}\propto \beta_A^2\,\beta_G^2$. As with the scalar DM case the constant of proportionality differs between the light and heavy cases.

We first compute the yield at $\Tc$ in the case that $m_j\ll T_{\rm max}$ obtaining for $\omega\neq3$
\begin{equation}
Y_{\rm light}(\Tc)\simeq\frac{90}{\pi^3\gss}\sqrt{\frac{10}{\gs}}\frac{\delta_j m^2M_{\rm Pl}}{\Lambda^4}\frac{1}{\omega-3} \Tc^{\frac{7-\omega}{1+\omega}} \left[\Tmax^{2\frac{\omega-3}{1+\omega}}-\Tc^{2\frac{\omega-3}{1+\omega}}\right].
\end{equation}
The corresponding boost factor for the fermion DM with  $\omega>3$ is given by
\beq
	B_\text{light}\simeq
		\frac{4}{3(\omega-3)}\left[\frac{\Tmax}{\Trh}\right]^{2\frac{\omega-3}{\omega+1}}.
\eeq

For fermion DM with $m_j\gg T$ the yield is given by (for $\omega\neq1/3$)
\begin{equation}
Y_{\rm heavy}(\Tc)\simeq\frac{90}{\pi^3\gss}\sqrt{\frac{10}{\gs}}
\frac{~\delta_j m^2 M_{\rm Pl}}{\Lambda^4 m_j^4}\frac{1}{3\omega-1}
\Tc^{\frac{7-\omega}{1+\omega}}
\left[\Tmax^{2\frac{3\omega-1}{1+\omega}}-\Tc^{2\frac{3\omega-1}{1+\omega}}\right],
\end{equation}
with and the boost factor for $\omega>1/3$ is 
\beq
	B_\text{heavy}\simeq
		\frac{20}{3(3\omega-1)}\left[\frac{\Tmax}{\Trh}\right]^{2\frac{3\omega-1}{\omega+1}}. 
\eeq
Observe, in particular, that the critical values for $\omega$ above which the boost factor can be significant is quite different in the fermion DM case to previous scenarios studied here.

\subsection{Higgs portals}
\label{3.4}

We highlight one last case, which is arguably the most natural way of coupling DM to the Standard Model, namely via the Higgs quadratic $|H|^2$. This operator  is the lowest order gauge and Lorentz invariant operator in the Standard Model and thus may be the most relevant when contracted with some operator $\mathcal{O}_{\rm DM}$ involving DM or hidden sector fields. The prospect of freeze-in via Higgs portals was recently studied in e.g.~\cite{Kolb:2017jvz,McDonald:2015ljz}. We are primarily interested in scenarios that provide modest (and unexpected) boosts to the relic density due to the cosmological evolution and, as can been seen in Figures~\ref{fig:boost0} and~\ref{fig:boost}, typically to arrive at a modest boost factor one requires the connector operator to be of at least mass dimension six. One simple example which presents itself is the case of vector DM $V_\mu$ which couples to the Higgs via the Lagrangian term $|H|^2V_{\mu\nu}V^{\mu\nu}$ where $V_{\mu\nu}=V_\mu V_\nu-V_\nu V_\mu$ denotes the corresponding field strength~\cite{Lebedev:2011iq}, this operator is mass dimension 6 thus corresponding to $n=2$ and with a boost factor which is parametrically
\beq
	B\simeq\begin{cases}
		\frac{2}{1-\omega} & \text{ for } \omega<1\,,\\[4pt]
		\frac83\frac{7-\omega}{(1+\omega)^2}\ln\frac{\Tmax}{\Trh} & \text{ for } \omega=1\,,\\[4pt]
		\frac{2}{\omega-1}\left[\frac{\Tmax}{\Trh}\right]^{4\frac{\omega-1}{1+\omega}} & \text{ for } \omega>1\,.
	\end{cases}
\eeq
In particular, we observe that in the case of $\omega=1$, which is the independently motivated scenario of kination domination, that the UV freeze-in abundance is logarithmically enhanced and that large boosts can occur in more non-standard scenarios with $\omega>1$. 

\vspace{-2mm}
\section{Concluding remarks}
\label{S4}\vspace{-2mm}

In the simple picture of DM the relic density is established during an era of radiation domination in which the evolution of the DM abundance is simple to track and is largely determined by the particle physics model, in particular the mass and couplings of the DM. However, if the DM abundance is generated via  UV freeze-in then it invariably occurs at the highest temperatures during the transition to radiation domination, and as such cosmology can potentially impact relic density calculations.  In this work we have highlighted that the abundance produced via UV freeze-in can be sensitive to the equation of state of the universe $\omega$ prior to the transition to radiation domination.

Specifically, we have demonstrated that for an initial equation of state $\omega$, UV freeze-in via an operator of mass dimension $5+n/2$ receives a parametric enhancement of $(T_{\rm max}/T_{\rm RH})^{n-n_c}$ for $n>n_c$ where the critical threshold is $n_c=2\times (3-\omega)/(1+\omega).$
For matter dominated initial state then $n_c=6$ and the details of the process of reheating are only important for operators with mass dimension 8 and higher, as such much of the existing literature on UV freeze-in will remain consistent with the implicit assumption that initially $\omega=0$.  However, the critical threshold above which the DM abundance is enhanced $n_c$ depends on $\omega$ and thus for non standard cosmologies the DM abundance can receive a significant boost even for portal operators with relatively low mass dimension. This is important for potential DM candidates that are typically under produced for instance gravitino in High Scale SUSY (as discussed) or the bino LSP in low scale SUSY.

Finally, we highlight that direct searches for DM produced via UV freeze-in (with or without a boost factor) is likely challenging outside of special constructions or benign corners of parameter space, however primordial gravitational wave production from the era prior to radiation domination could provide a potential probe in the future \cite{Assadullahi:2009nf,Alabidi:2013wtp, DEramo:2019tit, Bernal:2019lpc, Figueroa:2019paj}.

\acknowledgments
We thank Keith Olive and James Scargill for helpful discussions.
This project has received funding from the European Union's Horizon 2020 research and innovation programme under the Marie Sklodowska-Curie grant agreements 674896 and 690575.
NB is partially supported by Spanish MINECO under grant FPA2017-84543-P, from Universidad Antonio Nari\~no grants 2018204, 2019101 and 2019248, and by the ``Joint Excellence in Science and Humanities'' (JESH) program of the Austrian Academy of Sciences.
NB thanks the Erwin Schr\"odinger International Institute for hospitality while this work was completed.
FE is grateful to CERN theory group for their hospitality. 
JU is grateful for the hospitality and support of the Simons Center for Geometry and Physics (Program: Geometry \& Physics of Hitchin Systems) where some of this work was carried out, and acknowledges support from NSF grant DMS-1440140 while in residence at MSRI, Berkeley, CA during Fall 2019.

\bibliography{biblio}
\end{document}